\documentclass[aps]{revtex4}
\usepackage{amsfonts}
\usepackage{amsmath}
\usepackage{amssymb,epsf}

\begin{document}

\title{$P-V$ criticality and geometrothermodynamics of black holes \\
with Born-Infeld type nonlinear electrodynamics}
\author{S. H. Hendi$^{1,2}$\footnote{%
email address: hendi@shirazu.ac.ir}, S. Panahiyan$^{1}$\footnote{
email address: ziexify@gmail.com} and B. Eslam
Panah$^{1}$\footnote{ email address:
behzad$_{-}$eslampanah@yahoo.com}} \affiliation{$^1$ Physics
Department and Biruni Observatory, College of Sciences, Shiraz
University, Shiraz 71454, Iran\\
$^{2}$ Research Institute for Astronomy and Astrophysics of Maragha (RIAAM),
Maragha, Iran}

\begin{abstract}
In this paper, we take into account the black hole solutions of Einstein
gravity in the presence of logarithmic and exponential forms of nonlinear
electrodynamics. At first, we consider the cosmological constant as a
dynamical pressure to study the analogy of the black hole solutions with the
Van der Waals liquid--gas system in the extended phase space. We plot $P-v$,
$T-v$ and $G-T$ diagrams and investigate the phase transition of adS black
holes in the canonical ensemble. We study the nonlinearity effects of
electrodynamics and see how the power of nonlinearity affects critical
behavior and phase transition of the system. We also investigate the effects
of dimensionality on the critical values and analyze its crucial role.
Moreover, we show the changes in the universal ratio $P_{c}v_{c}/T_{c}$ for
variation of different parameters. In addition, we make a comparison between
linear and nonlinear electromagnetic fields and show that the lowest
critical temperature belongs to Maxwell theory. Also, we make some arguments
regarding to how power of nonlinearity brings the system to
Schwarzschild--like and Reissner--Nordstr\"{o}m--like limitations. Next, we
study the critical behavior of the system in context of heat capacity. We
show that critical behavior of system is similar to the one in phase
diagrams of extended phase space. We point out that phase transition points
of the extended phase space only appear as divergencies of heat capacity. We
also extend the study of phase transition points through
geometrothermodynamics (GTD) method. We introduce two new thermodynamical
metrics for extended phase space and show that divergencies of
thermodynamical Ricci scalar of the new metrics coincide with phase
transition points of the system. The characteristic behavior of these
divergencies, hence critical points is exactly the one that is obtained in
extended phase space and heat capacity. Then, we introduce a new method for
obtaining critical pressure and horizon radius by considering denominator of
the heat capacity. We show that there are several benefits that make this
approach favorable comparing to other ones.
\end{abstract}

\maketitle

\section{introduction}


In recent years, there has been an increasing interest in asymptotically
anti-de Sitter black holes. This is mainly based on the fact that there is
an equivalence of string theory on asymptotically anti de-Sitter spacetimes
and the quantum field theory living on the boundary of them \cite%
{AdS/CFT1,AdS/CFT2,AdS/CFT3,AdS/CFT4}. Although Einstein had inserted the
cosmological constant as a fixed parameter in the gravitational field
equations, in an increasing number of recent papers it was shown that it
might be regarded as a variable \cite{VariableLambda1,VariableLambda2}. In
other words, following the recent idea of including the cosmological
constant in the first law of black hole thermodynamics, the cosmological
constant is no longer a fixed parameter, but rather a thermodynamical
variable.

Regarding the extended phase space of black holes thermodynamics, one may
treat the cosmological constant as a dynamical pressure \cite%
{PressureLambda1,PressureLambda2,PressureLambda3,PressureLambda4,
PressureLambda5,PressureLambda6,PressureLambda7,PressureLambda8,PressureLambda9, PressureLambda10,PressureLambda11}%
. The results are much richer thermodynamics than heretofore, and
interestingly, it is shown that the critical behavior of black holes is
analogous to the Van der Waals liquid-gas phase transition. So,
investigation of the mentioned extending phase space and its phase
transition have gained a lot of attention recently \cite%
{PressureLambda1,PressureLambda2,PressureLambda3,PressureLambda4,
PressureLambda5,PressureLambda6,PressureLambda7,PressureLambda8,PressureLambda9, PressureLambda10,PressureLambda11}%
. Thermodynamic properties and phase transition of the asymptotically anti
de-Sitter black holes was studied by Hawking and Page \cite{HawkingPage}.
Witten \cite{Witten} reconsidered the Hawking-Page phase transition in the
context of gauge theory and AdS/CFT correspondence. Generally we should note
that phase transition plays an important role in describing different
phenomena in thermodynamics and quantum point of views.


On the other side, although Maxwell theory is capable of describing
different phenomena in electrodynamics domain, it fails regarding some
important issues (for various limitations of the Maxwell theory see Ref.
\cite{delph1,delph2} for more details). In order to solve these problems,
one may regard the nonlinear electrodynamics \cite%
{Hassaine1,Hassaine2,Hassaine3,Hassaine4,Hassaine5,Hassaine6,Oliveira1,Oliveira2,Soleng,HendiJHEP}%
. In recent years, investigation of nonlinear electrodynamics has got a new
impetus. Strong motivation comes from developments in string/M-theory, which
is a promising approach to quantum gravity \cite{Seiberg}. It has been shown
that the Born-Infeld \cite{Born-Lnfeld} (BI) type theories are specific in
the context of NLED models and naturally arise in the low-energy limit of
heterotic string theory \cite%
{Fradkin1,Fradkin2,Fradkin3,Fradkin4,Fradkin5,Fradkin6}. In recent years,
other BI types of NLED have been introduced, in which can remove the
divergency of the electric field of point charge near the origin. The
Lagrangians of logarithmic and exponential forms of NLED theories were,
respectively, proposed by Soleng \cite{Soleng} and Hendi \cite{HendiJHEP}
with the following explicit forms
\begin{equation}
\mathcal{L(F)}=\left\{
\begin{array}{c}
\beta ^{2}\left( \exp (-\frac{\mathcal{F}}{\beta ^{2}})-1\right) ,\;\;\;{ENEF%
} \vspace{0.2cm} \\
-8\beta ^{2}\ln \left( 1+\frac{\mathcal{F}}{8\beta ^{2}}\right) ,\;\;\;{LNEF}%
\end{array}%
\right. .  \label{NLED}
\end{equation}
where $\beta$ is nonlinearity parameter.

In this paper, we consider asymptotically anti-de Sitter black hole
solutions of the Einstein gravity in the presence of the recent BI type NLED
to investigate the extended phase space thermodynamics and $P-v$ criticality
of the solutions and also phase transition points through GTD.

Thermodynamical geometry is another approach for studying phase transition
of black holes. It means that, by studying the divergence points of
thermodynamical Ricci scalar (TRS), we can investigate phase transition
points for black holes. In other words, it is expected that divergencies of
TRS coincide with phase transition points of the black holes. In 1975
Weinhold \cite{Weinhold} introduced differential geometric concepts into
ordinary thermodynamics by considering a kind of metric defined as the
second derivatives of internal energy with respect to entropy and other
extensive quantities for a thermodynamical system. After that, Ruppeiner
\cite{Ruppeiner} introduced another metric and defined the minus second
derivatives of entropy with respect to the internal energy and other
extensive quantities. It is notable that, the Ruppeiner metric is conformal
to the Weinhold metric with the inverse temperature as the conformal factor.
Both thermodynamical metrics have been applied to study the thermodynamical
geometry of ordinary systems \cite{Janyszek1986,Brody,Dolan2002,Janke2004}.
In particular, it was found that the Ruppeiner geometry carries information
regarding phase structure of thermodynamical systems. Because of their
success for their applications in ordinary thermodynamical systems, they
have also been employed to study black hole phase structures which led to
interesting results \cite{Ferrara,Cai1999,Aman2003,Carlip,Mirza}.

Since these two approaches fail in order to describe phase transition of
several black holes \cite{HPEM}, Quevedo proposed new types of
thermodynamical metrics for studying geometrical structure of the black hole
thermodynamics \cite{Quevedo2007,Quevedo2008}. This method was employed to
study the geometrical structure of the phase transition of black holes \cite%
{Mo,Zhang,QuevedoP2011,Han,Bravetti,HendiAnn2} and proved to be a strong
machinery for describing phase transition of black holes. However, this
approach was not without any problem \cite{HPEM}. Hence for eliminating the
problems of previous thermodynamical metrics, Hendi et al. proposed a new
metric in Ref. \cite{HPEM}.

Let us begin with the following $d$-dimensional spherically symmetric line
element
\begin{equation}
ds^{2}=-g(r)dt^{2}+\frac{dr^{2}}{g(r)}+r^{2}d\Omega _{d-2}^{2},
\label{Metric}
\end{equation}%
where $d\Omega _{d}^{2}$ denotes the standard metric of $d$-dimensional
sphere $S^{d}$ with the volume $\omega _{d}$. In what follows, we consider
Einstein gravity coupled with the mentioned BI type NLED models \cite%
{HendiAnn1,HendiAnn2,Hendiothers1,Hendiothers2}. It was shown that,
regardless of gravitational sector, one may use the following
electromagnetic field equation to obtain related gauge potential
\begin{equation}
\nabla _{\mu }\left( \frac{d\mathcal{L(F)}}{d\mathcal{F}}F^{\mu \nu }\right)
=0,  \label{MaxwellEq}
\end{equation}%
in which the nonzero component of gauge potential is $A_{t}$ \cite%
{HendiAnn1,HendiAnn2}
\begin{equation}
A_{t}=\left\{
\begin{array}{ll}
\frac{{\beta {r}\sqrt{L{w}}}}{{2(d-3)(3d-7)}}\left[ {3d-7+(d-2)\varpi L}%
_{W}{}\right] , & \;{ENEF}\vspace{0.2cm} \\
\frac{{2{\beta ^{2}}{r}^{d-1}}}{{q}\left( d-1\right) }\left[ {1-{\varpi }%
^{\prime }}\right] , & \;{LNEF}%
\end{array}%
\right. ,  \label{At}
\end{equation}%
where%
\begin{eqnarray*}
{\varpi } &{=}&_{1}{F_{1}}\left( {[1],\,\left[ {\frac{{5d-11}}{{2d-4}}}%
\right] ,\,}\frac{{L}_{W}}{{2d-4}}\right) , \\
{\varpi }^{\prime } &{=}&_{2}{F_{1}}\left( {\left[ {\frac{-1}{2},\,\frac{{1-d%
}}{{2d-4}}}\right] ,\,\left[ {\frac{d{-3}}{{2d-4}}}\right] ,}1-\Gamma
^{2}\right) ,
\end{eqnarray*}%
and the Lambert function $L_{W}=LambertW\left( \frac{4q^{2}}{\beta
^{2}r^{2d-4}}\right) $, $\Gamma =\sqrt{1+\frac{q^{2}}{\beta ^{2}r^{2d-4}}}$
and $q$ is an integration constant related to the the total electric charge $%
Q=\frac{V_{d-2}}{8\pi }q$. The electric potential $U$, measured at infinity
with respect to the event horizon, is $U=-A_{t}$. It is easy to show that
the electric field may be written as \cite{HendiAnn1,HendiAnn2}
\begin{equation}
E(r)=F_{tr}=\frac{Q}{r^{2}}\times \left\{
\begin{array}{ll}
e^{-\frac{L_{W}}{2}}, & \;{ENEF} \\
\frac{2}{\Gamma +1}, & \;{LNEF}%
\end{array}%
\right. .  \label{E}
\end{equation}

\section{Extended phase space and $P-V$ criticality in Einstein gravity}

Starting with the Einstein gravity in the presence of NLED, we consider the
following field equation
\begin{equation}
G_{\mu \nu }+\Lambda g_{\mu \nu }=\frac{1}{2}g_{\mu \nu }\mathcal{L(F)}-2%
\frac{d\mathcal{L(F)}}{d\mathcal{F}}F_{\mu \lambda }F_{\nu }^{\lambda },
\label{EinsteinEq}
\end{equation}%
with the following solutions for the metric function \cite%
{HendiAnn1,HendiAnn2}
\begin{equation}
g(r)=1-\frac{2\Lambda r^{2}}{\left( d-1\right) (d-2)}-\frac{m}{r^{d-3}}-%
\frac{8\beta ^{2}r^{2}\Upsilon }{\left( d-1\right) \left( d-2\right) },
\end{equation}
\begin{eqnarray*}
\Upsilon &=&\left\{
\begin{array}{c}
1+\frac{2(d-1)q}{\beta r^{d-1}}\left[ \int \left( \sqrt{L_{W}}-\frac{1}{%
\sqrt{L_{W}}}\right) dr\right] ,\;{ENEF} \vspace{0.5cm} \\
\frac{\left( 2d-3\right) (\Gamma -1)}{\left( d-1\right) }-\ln \left( \frac{%
1+\Gamma }{2}\right) +\mathcal{H},\;\;\;\;\; \; \; \;\;\; {LNEF}%
\end{array}
\right. , \\
\mathcal{H} &=&\frac{\left( d-2\right) ^{2}\left( 1-\Gamma ^{2}\right){%
_{2}F_{1}}\left( \left[ \frac{1}{2},\frac{d-3}{2d-4}\right] ,\left[ \frac{%
3d-7}{2d-4}\right] ,1-\Gamma ^{2}\right) }{\left( d-1\right) (d-3)},
\end{eqnarray*}%
where $m$ is an integration constant which is related to the total mass.
Looking for the curvature singularity of the metric (\ref{Metric}), one
finds that the Kretschmann scalar diverges at $r=0$. In addition, numerical
analysis shows that the metric functions have at least one real positive
root. Moreover, using series expansion of metric function for large values
of $r$, we find that the dominant term is related to $\Lambda $.
Consequently, we deduce that these solutions may be interpreted as
asymptotically anti de-Sitter black holes.

Now, we take into account the surface gravity interpretation to obtain the
Hawking temperature of the mentioned black hole solutions
\begin{eqnarray}
T &=&\frac{-2\Lambda r_{+}^{2}+\left( d-2\right) \left( d-3\right) -\Phi }{%
4\pi r_{+}\left( d-2\right) }, \vspace{0.5cm}  \label{Temp1} \\
\Phi &=&\left\{
\begin{array}{c}
\beta ^{2}r_{+}^{2}\left( \left[ 1+\left( \frac{2E}{\beta }\right) ^{2}%
\right] e^{\frac{-2E^{2}}{\beta ^{2}}}-1\right) ,\;\;\;{ENEF} \vspace{0.3cm}
\\
\;8r_{+}^{2}\beta ^{2}\ln \left[ 1-\left( \frac{E}{2\beta }\right) ^{2}%
\right] +\frac{4r_{+}^{2}E^{2}}{1-\left( \frac{E}{2\beta }\right) ^{2}}%
,\;\;\;{LNEF}%
\end{array}%
\right.  \notag
\end{eqnarray}

The finite mass of black hole can be obtained by using the behavior of the
metric at large distances \cite{Brewin}
\begin{equation}
M=\frac{\omega _{d-2}\left( d-2\right) m}{16\pi },  \label{Mass}
\end{equation}%
where one may obtain the parameter $m$ from the fact that the metric
functions vanish at the event horizon, $r_{+}$. The black hole entropy of
Einstein gravity may be determined from the area law
\begin{equation}
S=\frac{\omega _{d-2}}{4}r_{+}^{d-2}.  \label{EntropyE}
\end{equation}

Here, we regard $\Lambda $ as a thermodynamical pressure $P=\frac{-\Lambda }{%
8\pi }$ and its corresponding conjugate quantity is the thermodynamical
volume which one can obtain with following relation

\begin{equation}
V=\left( \frac{\partial H}{\partial P}\right) _{S,Q}.  \label{Volume}
\end{equation}

Due to fact that we are considering cosmological constant as thermodynamical
pressure, the interpretation of mass will be different from previous one.
Usually mass of the black hole is interpreted as internal energy. But by
considering such relation between cosmological constant and thermodynamical
pressure, the interpretation of mass will become enthalpy. In other words,
mass of the black hole has more contribution to thermodynamical construction
of the system. As a result of this new insight regarding mass, Gibbs free
energy of the system will be given by
\begin{equation}
G=H-TS=M-TS
\end{equation}

Now, we would like to study the phase transition of Einstein black hole
solutions in canonical ensemble with the mentioned NLED models. Using Eq. (%
\ref{Temp1}), one can obtain the following equation of state
\begin{equation}
P=\frac{\left( d-2\right) T}{4r_{+}}-\frac{\left( d-2\right) \left(
d-3\right)-\Phi }{16\pi r_{+}^{2} },  \label{Pressure}
\end{equation}
where $r_{+}$ is linear function of the specific volume $v$ in geometric
unit \cite{PressureLambda1,PressureLambda2,PressureLambda3,PressureLambda4,
PressureLambda5,PressureLambda6,PressureLambda7,PressureLambda8,PressureLambda9, PressureLambda10,PressureLambda11}%
. In general, for these thermodynamical systems, we have the following
result for volume

\begin{equation}
V=\left( \frac{\partial M}{\partial P}\right) _{S,Q}=\frac{\omega _{d-2}{%
r_{+}}^{d-1}}{d-1},  \label{V}
\end{equation}%
which is in agreement with the topological structure of our spacetime
(spherical symmetric). This approval is a confirmation for considering
cosmological constant as a thermodynamical pressure.

Besides, we know that the Smarr formula may be extended to nonlinear
theories of electrodynamics \cite{SmarrNew1,SmarrNew2,SmarrNew3,SmarrNew4}.
In order to obtain an extension of the first law and its related modified
Smarr formula, one can use geometrical techniques (scaling argument). Here, $%
M$ should be the function of entropy, pressure, charge and BI parameter \cite%
{SmarrNew1,SmarrNew2,SmarrNew3,SmarrNew4}. Regarding obtained quantities, we
find that they satisfy the following differential form
\begin{equation}
dM=TdS+\Phi dQ+VdP+\mathcal{B}d\beta .  \label{GenFirstLaw}
\end{equation}
where we have calculated $T$ and $\Phi $, and one can obtain
\begin{eqnarray*}
V &=&\left( \frac{\partial M}{\partial P}\right) _{S,Q,\beta }, \\
\mathcal{B} &=&\left( \frac{\partial M}{\partial \beta }\right) _{S,Q,P}.
\end{eqnarray*}

In addition, taking into account the scaling argument, we can obtain the
generalized Smarr relation (per unit volume $\omega _{d-2}$) for our
asymptotically adS solutions in the extended phase space
\begin{equation}
(d-3)M=(d-2)TS+(d-3)Q\Phi -2PV-\mathcal{B}\beta  \label{Smarr2}
\end{equation}%
where
\begin{eqnarray*}
V &=&\frac{r_{+}^{d-1}}{d-1}, \\
\left. \mathcal{B}\right\vert _{\text{ENED}} &=&\frac{q(d-2)r_{+}F\left(
[1],[\frac{5d-11}{2d-4}],\frac{L_{W+}}{2d-4}\right) }{8\pi (d-1)(3d-7)\left(
L_{W+}\right) ^{\frac{-3}{2}}}-\frac{\beta r_{+}^{d-1}}{8\pi (d-1)}+\frac{%
q\beta r_{+}^{d}\sqrt{L_{W+}}\left( 1-L_{W+}\right) }{8\pi (d-1)\left(
1+L_{W+}\right) }+\frac{2qr_{+}\left( 1+L_{W+}\right) ^{-1}}{8\pi (d-1)\sqrt{%
L_{W+}}}, \\
\left. \mathcal{B}\right\vert _{\text{LNED}} &=&\frac{\beta r_{+}^{d-1}}{%
2\pi }\left[ \frac{(d-2)\left( \Gamma _{+}^{2}-1\right) }{(d-1)^{2}}F\left( %
\left[ \frac{1}{2},\frac{d-3}{2d-4}\right] ,\left[ \frac{3d-7}{2d-4}\right]
,1-\Gamma _{+}^{2}\right) +\frac{2\ln \left( \frac{1+\Gamma _{+}}{2}\right)
}{(d-1)}+\frac{(3d-5)\left( 1-\Gamma _{+}\right) }{(d-1)^{2}}\right] .
\end{eqnarray*}


The next step will be devoted to obtain critical values. In order to
investigate phase transition and the behavior of these thermodynamical
systems, we work in geometric unit and draw graphs of $P-v$, $T-v$ and $G-T$
diagrams.

\begin{figure}[tbp]
$%
\begin{array}{ccc}
\epsfxsize=5cm \epsffile{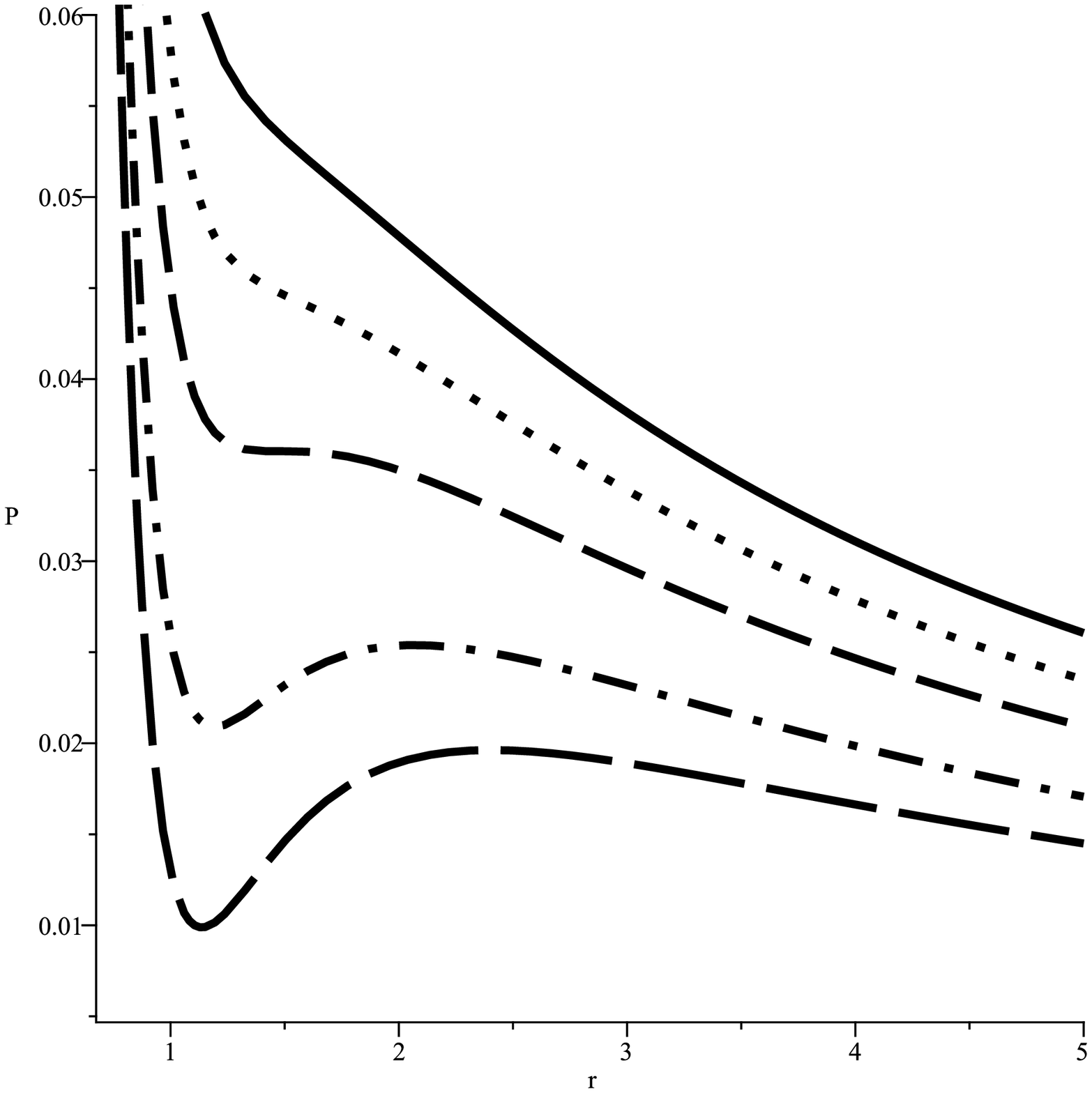} & \epsfxsize=5cm %
\epsffile{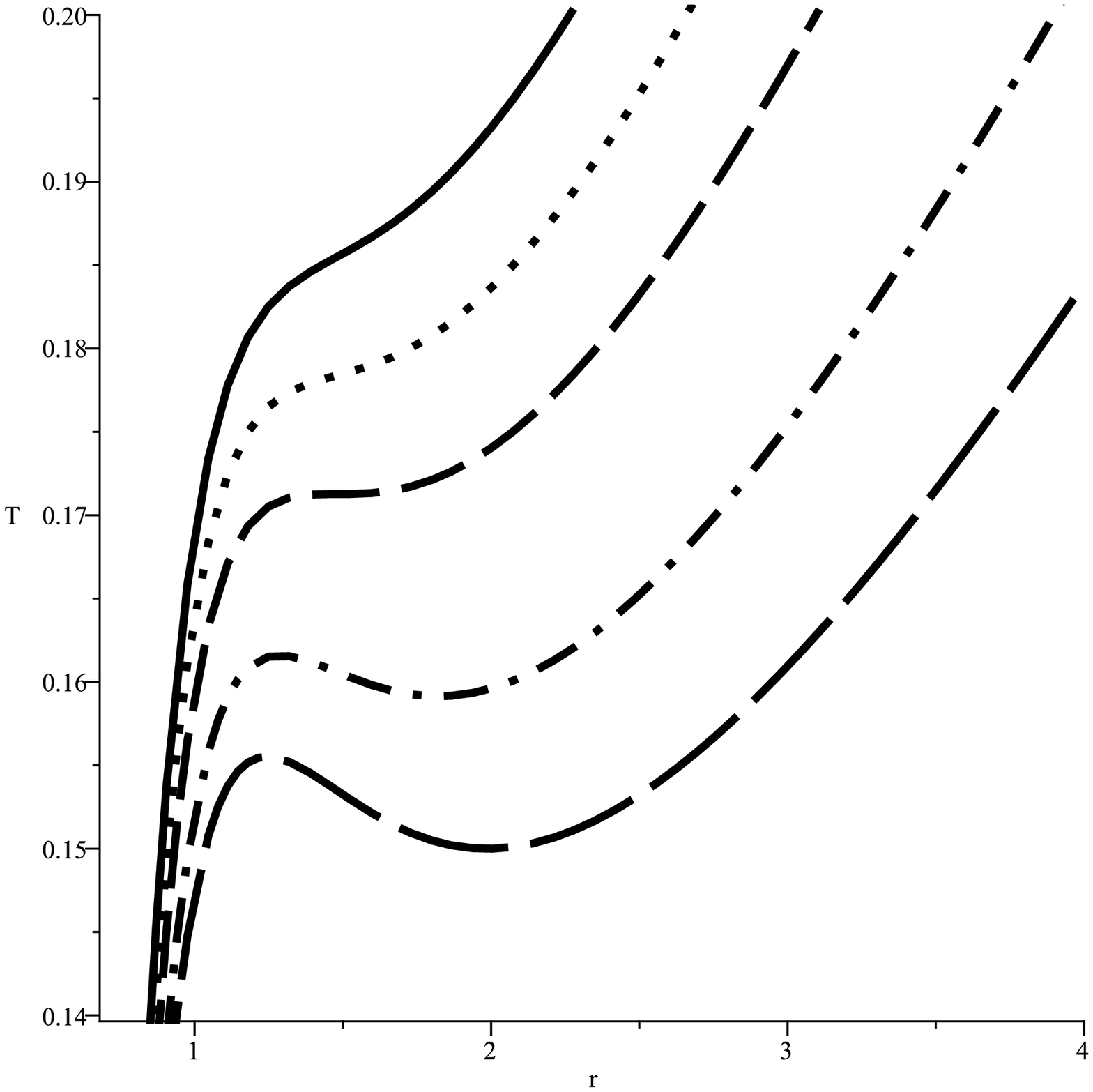} & \epsfxsize=5cm \epsffile{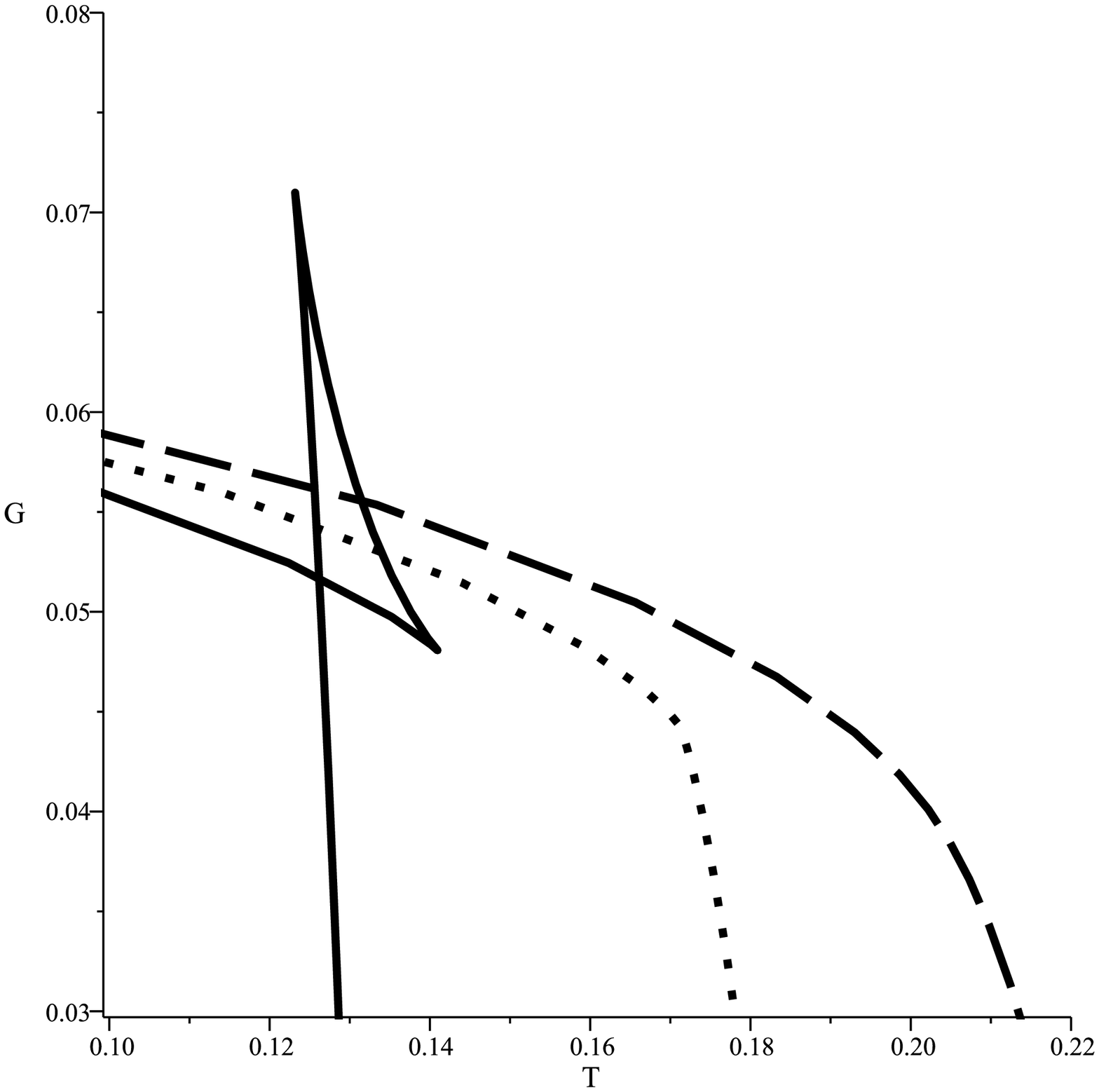}%
\end{array}
$%
\caption{\textbf{"LNEF branch:"} $P-v$ (left), $T-v$ (middle) and $G-T$
(right) diagrams for $q=1$,$\protect\beta=1$ and $d=5$. \newline
$P-v$ diagram, from up to bottom $T=1.2T_{c}$, $T=1.1T_{c}$, $T=T_{c}$, $%
T=0.85T_{c}$ and $T=0.75T_{c}$, respectively. \newline
$T-v$ diagram, from up to bottom $P=1.2P_{c}$, $P=1.1P_{c}$, $P=P_{c}$, $%
P=0.85P_{c}$ and $P=0.75P_{c}$, respectively. \newline
$G-T$ diagram, for $P=0.5P_{c}$ (continuous line), $P=P_{c}$ (dot line) and $%
P=1.5P_{c}$ (dashed line).}
\label{Fig2Ein}
\end{figure}

\begin{figure}[tbp]
$%
\begin{array}{ccc}
\epsfxsize=5cm \epsffile{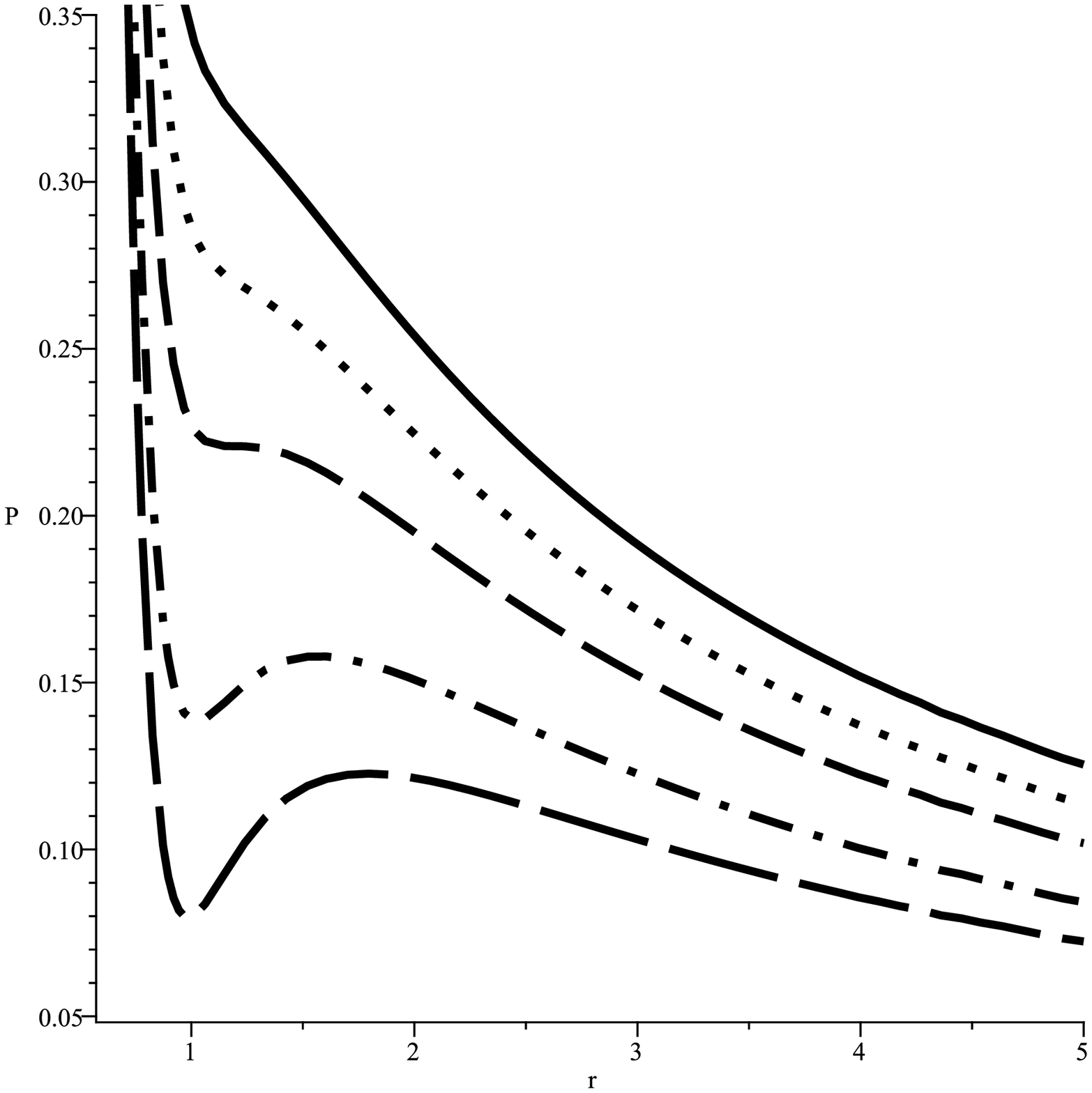} & \epsfxsize=5cm %
\epsffile{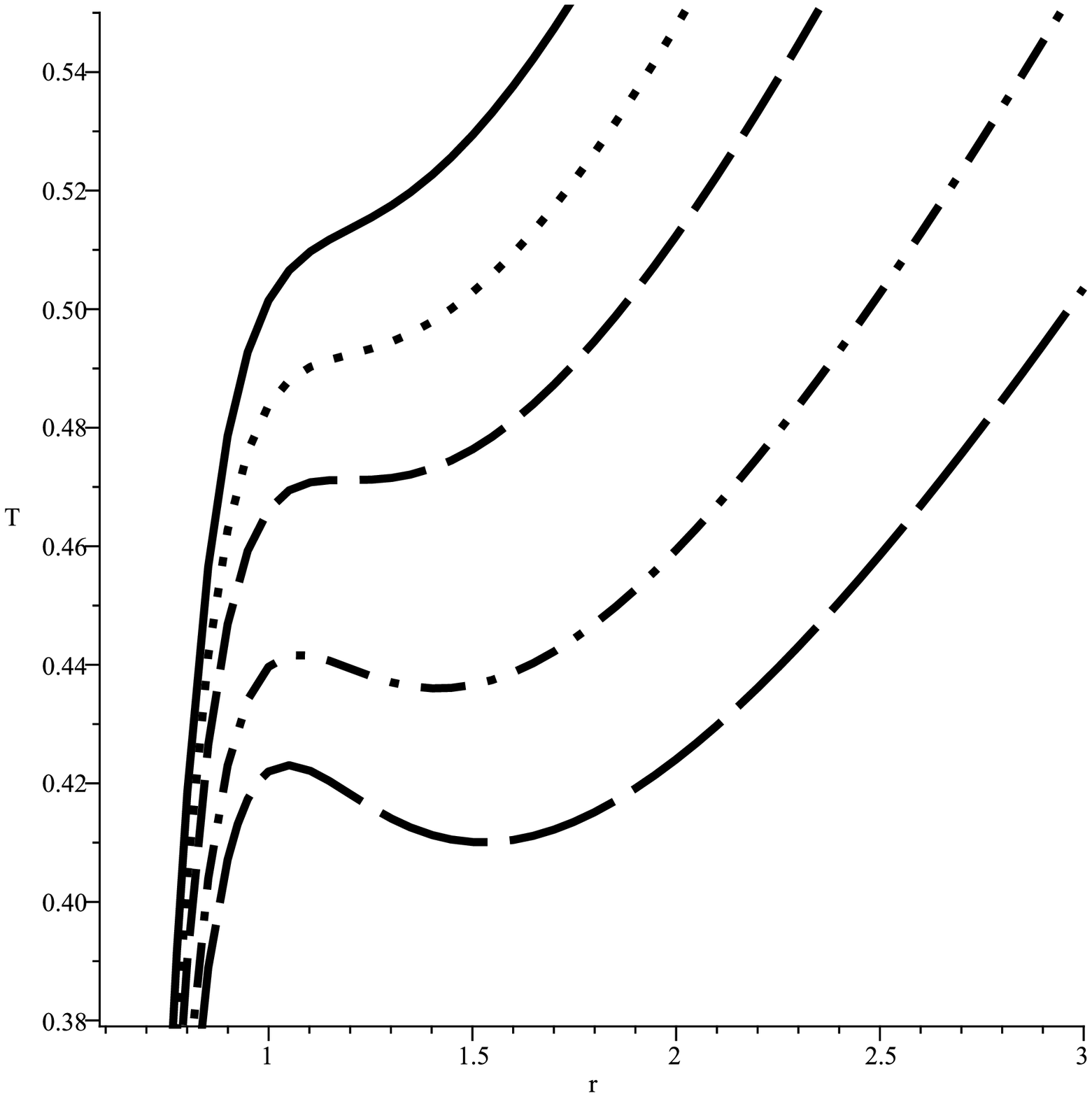} & \epsfxsize=5cm \epsffile{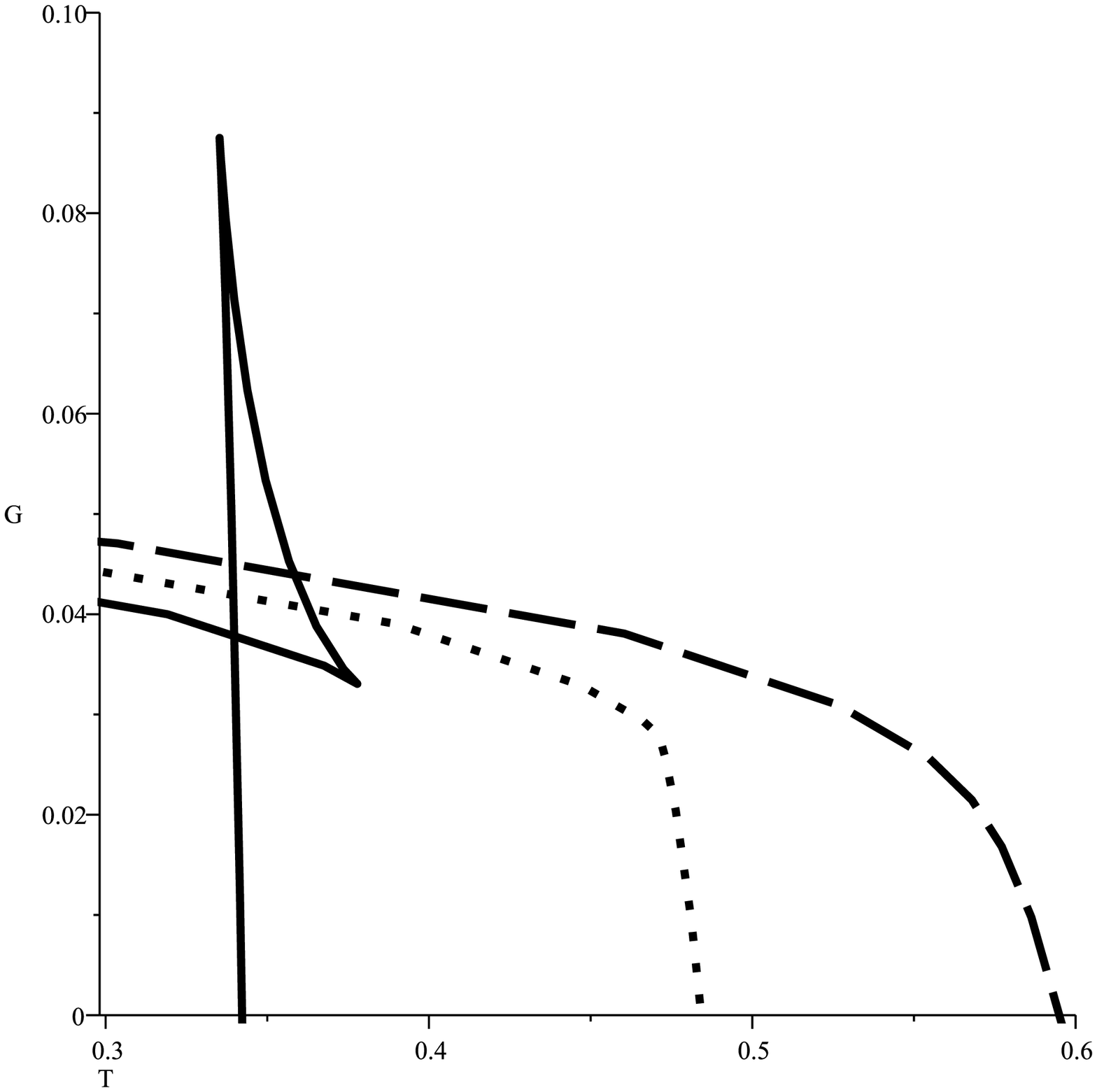}%
\end{array}
$%
\caption{\textbf{"LNEF branch:"} $P-v$ (left), $T-v$ (middle) and $G-T$
(right) diagrams for $q=1$, $\protect\beta=1$ and $d=7$. \newline
$P-v$ diagram, from up to bottom $T=1.2T_{c}$, $T=1.1T_{c}$, $T=T_{c}$, $%
T=0.85T_{c}$ and $T=0.75T_{c}$, respectively. \newline
$T-v$ diagram, from up to bottom $P=1.2P_{c}$, $P=1.1P_{c}$, $P=P_{c}$, $%
P=0.85P_{c}$ and $P=0.75P_{c}$, respectively. \newline
$G-T$ diagram, for $P=0.5P_{c}$ (continuous line), $P=P_{c}$ (dot line) and $%
P=1.5P_{c}$ (dashed line).}
\label{Fig5Ein}
\end{figure}

\begin{figure}[tbp]
$%
\begin{array}{ccc}
\epsfxsize=5cm \epsffile{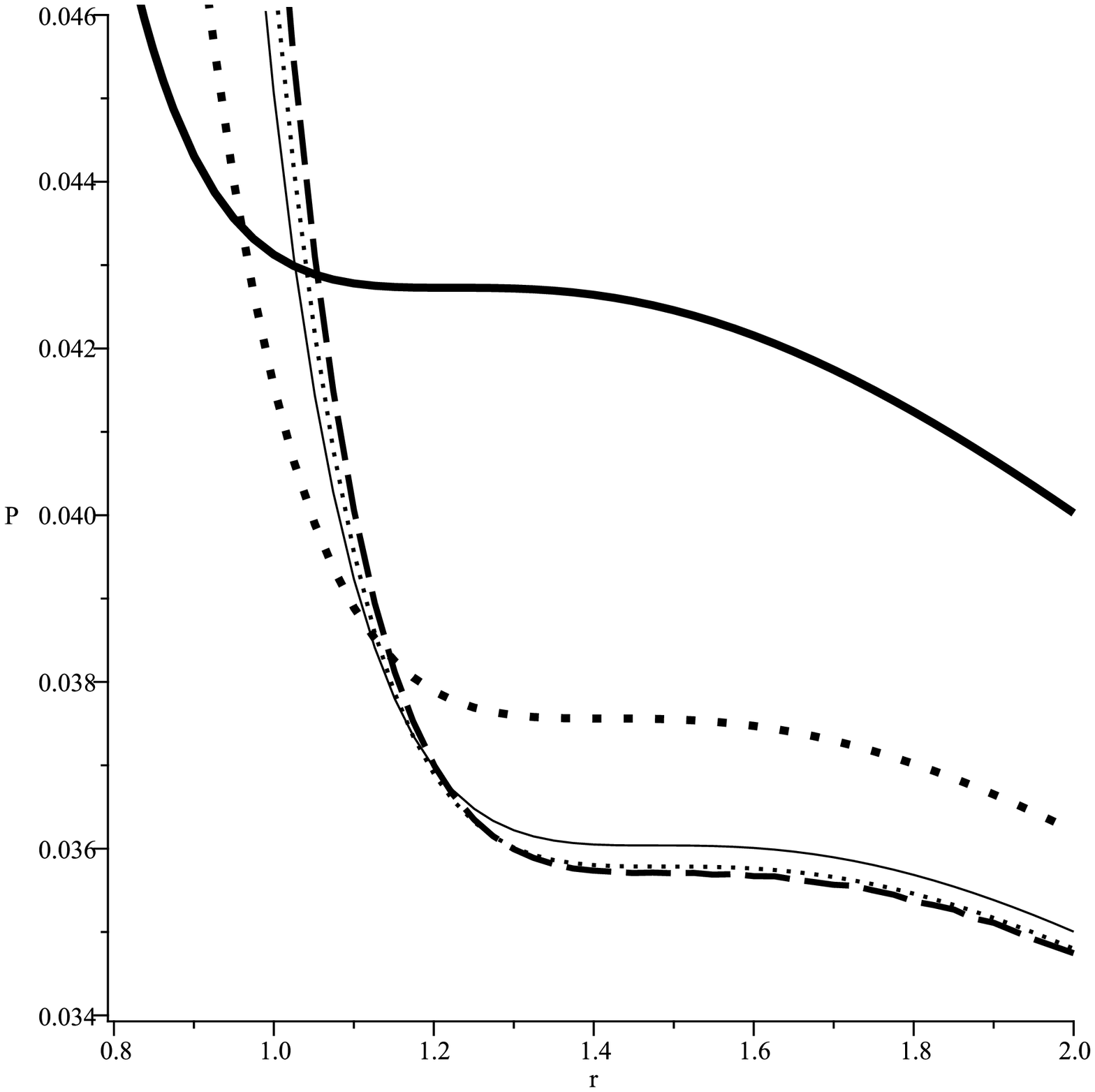} & \epsfxsize=5cm %
\epsffile{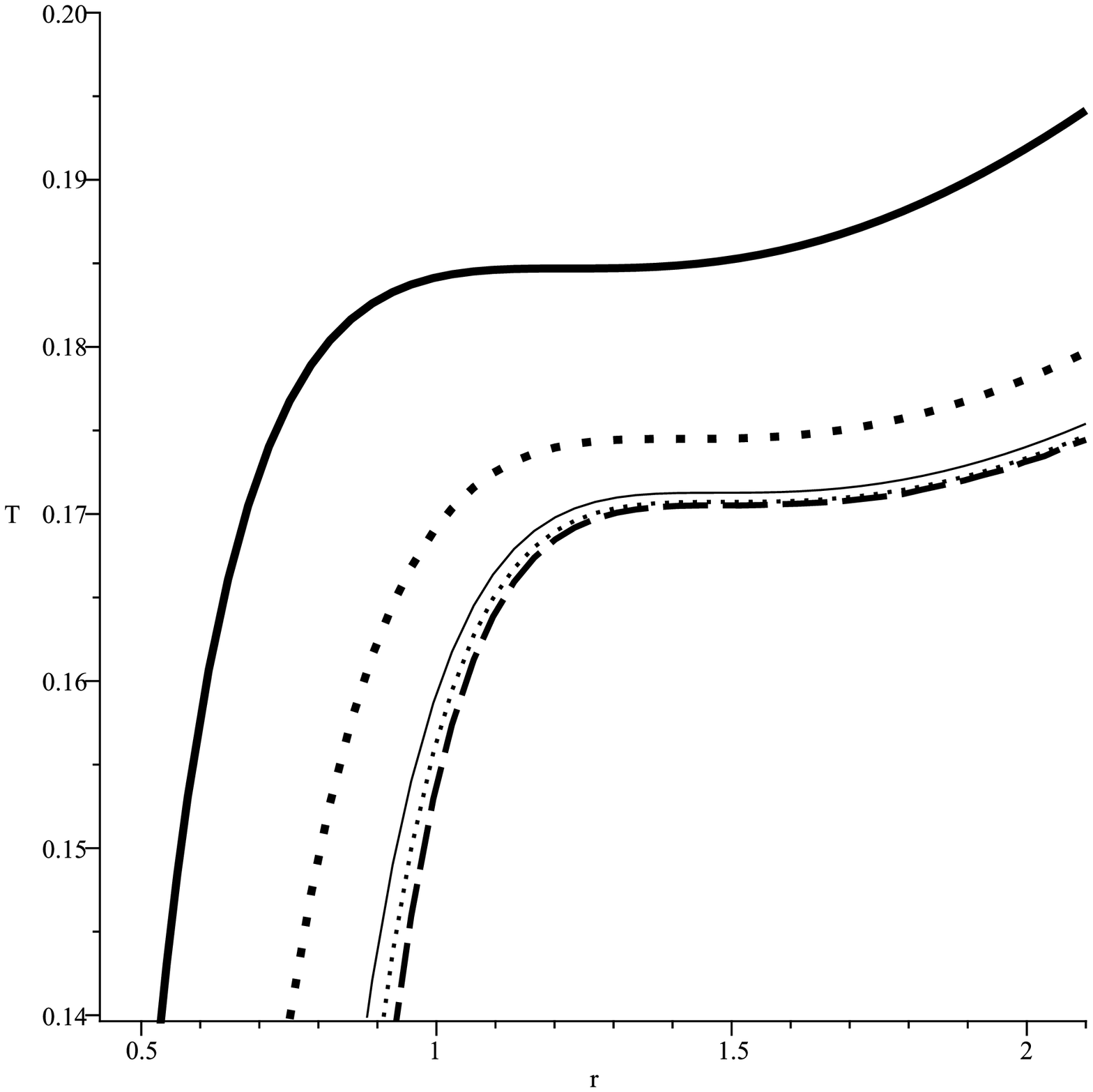} & \epsfxsize=5cm %
\epsffile{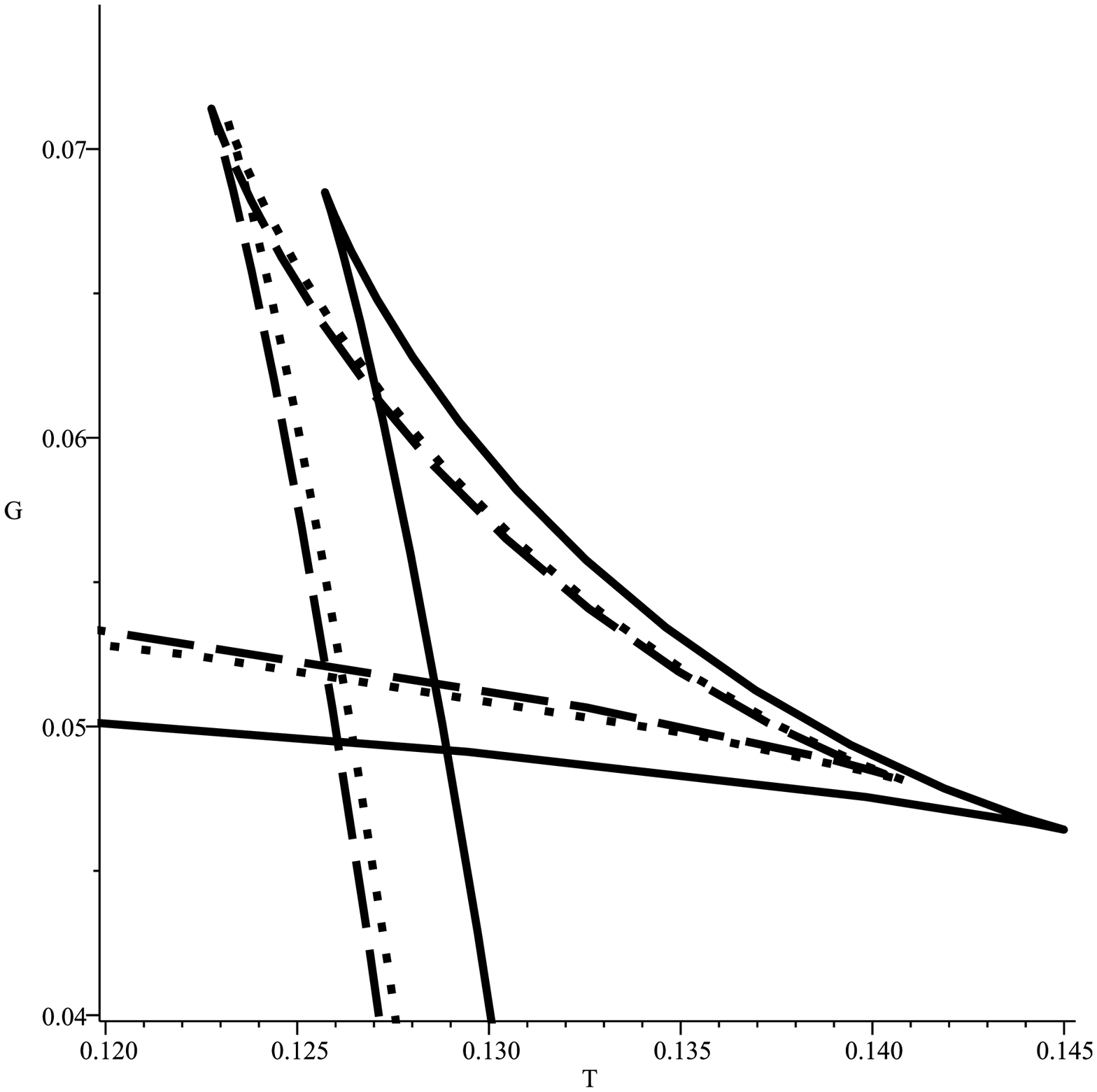}%
\end{array}
$%
\caption{\textbf{"LNEF branch:"} $P-v$ (left), $T-v$ (middle) and $G-T$
(right) diagrams for $q=1$ and $d=5$. \newline
$P-v$ diagram, for $T=T_{c}$, $\protect\beta =0.3$ (bold line), $\protect%
\beta =0.5$ (bold dot line), $\protect\beta =1$ (continuous line), $\protect%
\beta =1.5$, (dot line) and $\protect\beta =10$ (dash line). \newline
$T-v$ diagram, for $P=P_{c}$, $\protect\beta =0.3$ (bold line), $\protect%
\beta =0.5$ (bold dot line), $\protect\beta =1$ (continuous line), $\protect%
\beta =1.5$, (dot line) and $\protect\beta =10$ (dash line). \newline
$G-T$ diagram, for $P=0.5P_{c}$, $\protect\beta =0.5$ (continuous line), $%
\protect\beta =1$, (dot line) and $\protect\beta =1.5$ (dash line).}
\label{Fig7Ein}
\end{figure}

\begin{figure}[tbp]
$%
\begin{array}{ccc}
\epsfxsize=5cm \epsffile{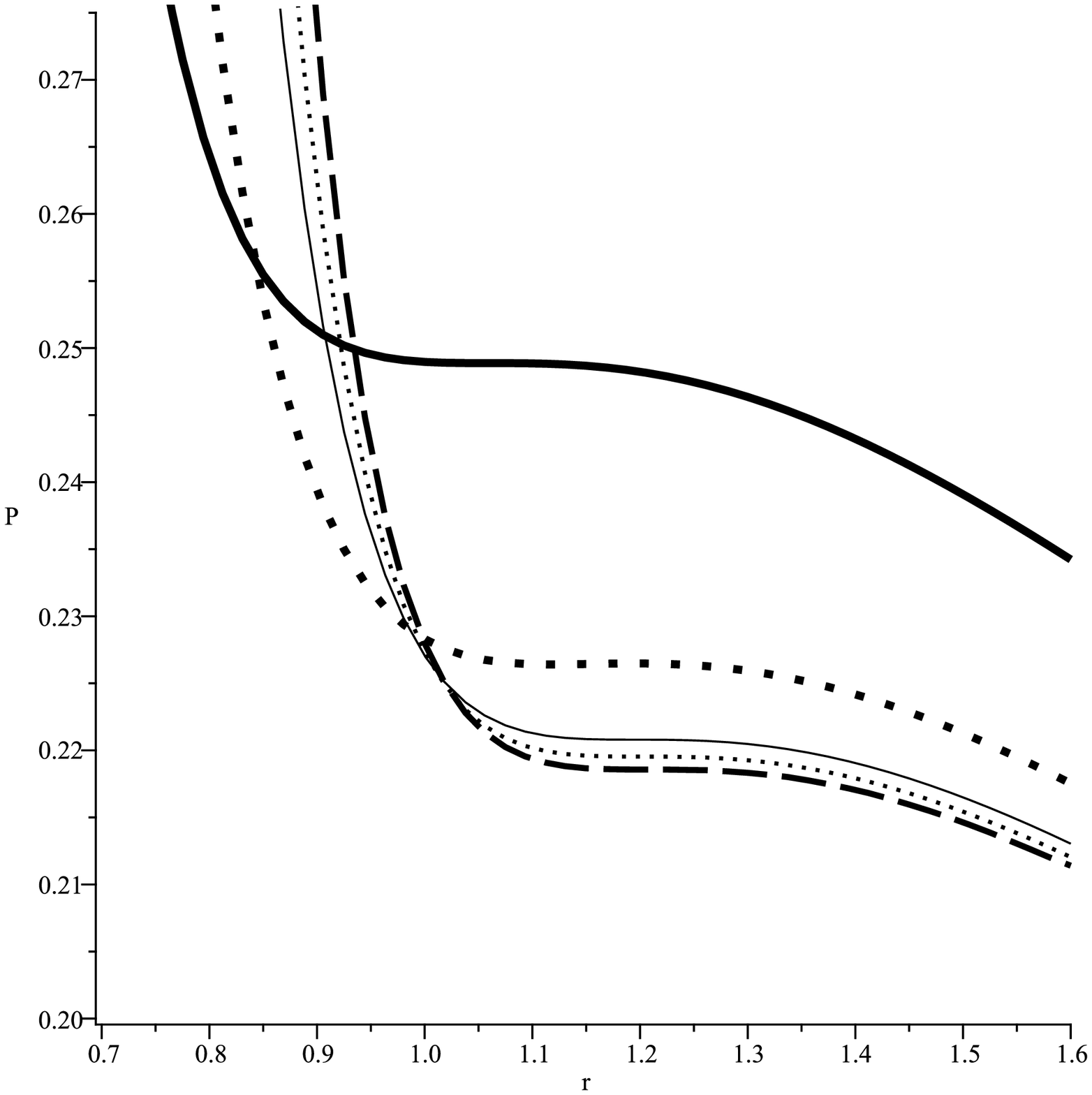} & \epsfxsize=5cm %
\epsffile{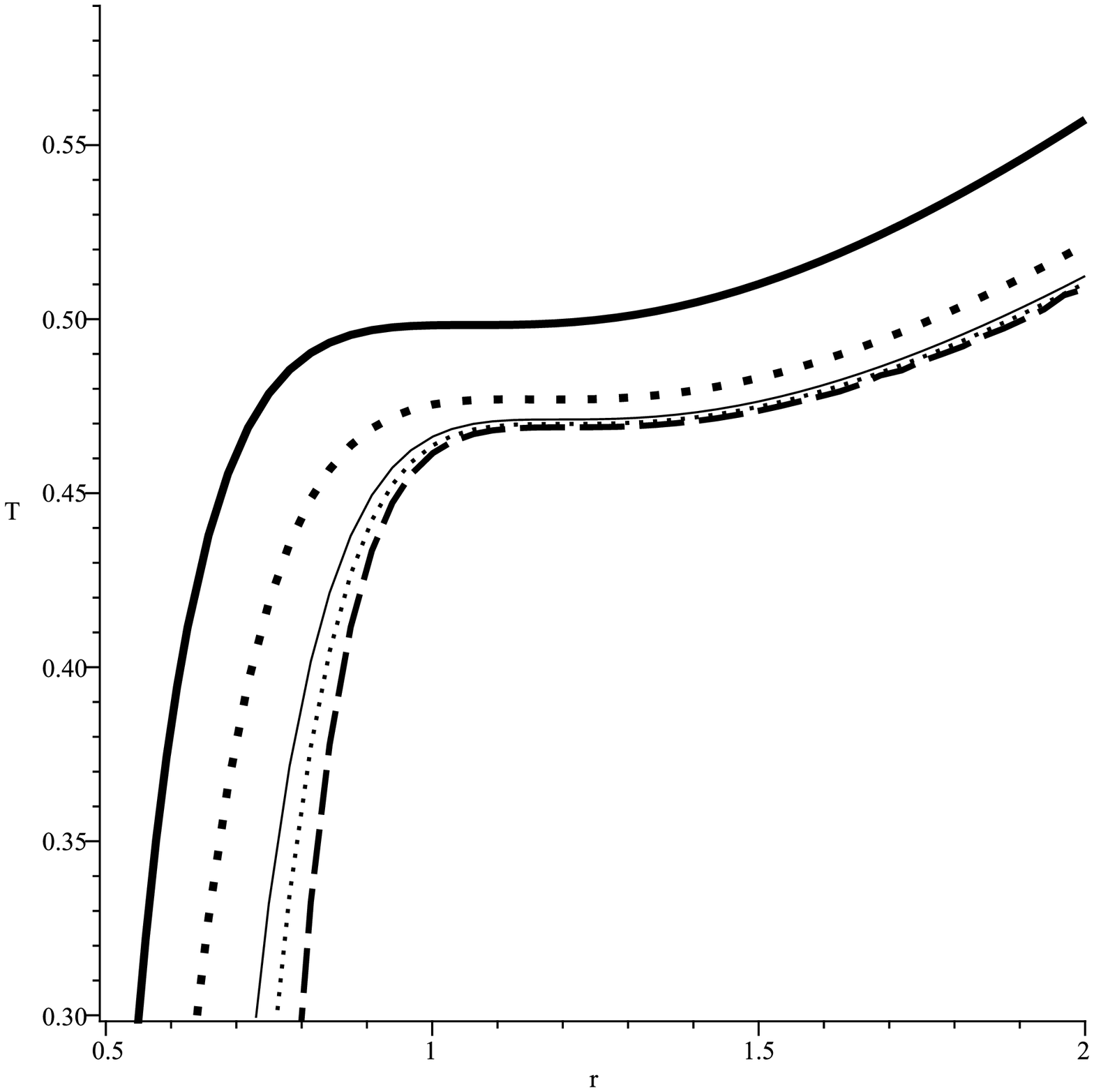} & \epsfxsize=5cm %
\epsffile{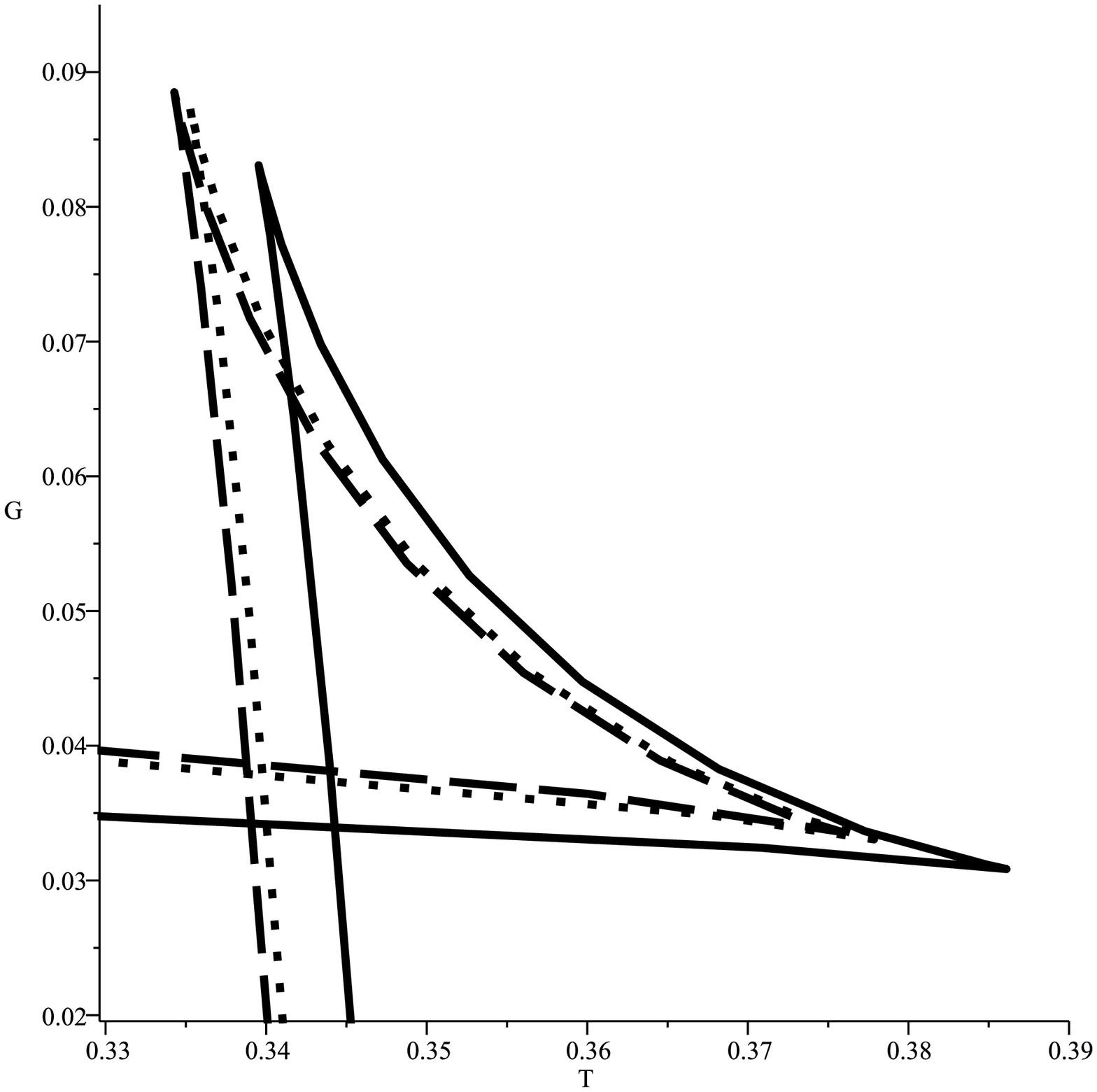}%
\end{array}
$%
\caption{\textbf{"LNEF branch:"} $P-v$ (left), $T-v$ (middle) and $G-T$
(right) diagrams for $q=1$ and $d=7$. \newline
$P-v$ diagram, for $T=T_{c}$, $\protect\beta =0.3$ (bold line), $\protect%
\beta =0.5$ (bold dot line), $\protect\beta =1$ (continuo us line), $\protect%
\beta =1.5$, (dot line) and $\protect\beta =10$ (dash line). \newline
$T-v$ diagram, for $P=P_{c}$, $\protect\beta =0.3$ (bold line), $\protect%
\beta =0.5$ (bold dot line), $\protect\beta =1$ (continuous line), $\protect%
\beta =1.5$, (dot line) and $\protect\beta =10$ (dash line). \newline
$G-T$ diagram, for $P=0.5P_{c}$, $\protect\beta =0.5$ (continuous line), $%
\protect\beta =1$, (dot line) and $\protect\beta =1.5$ (dash line). }
\label{Fig8Ein}
\end{figure}

\begin{figure}[tbp]
$%
\begin{array}{ccc}
\epsfxsize=5cm \epsffile{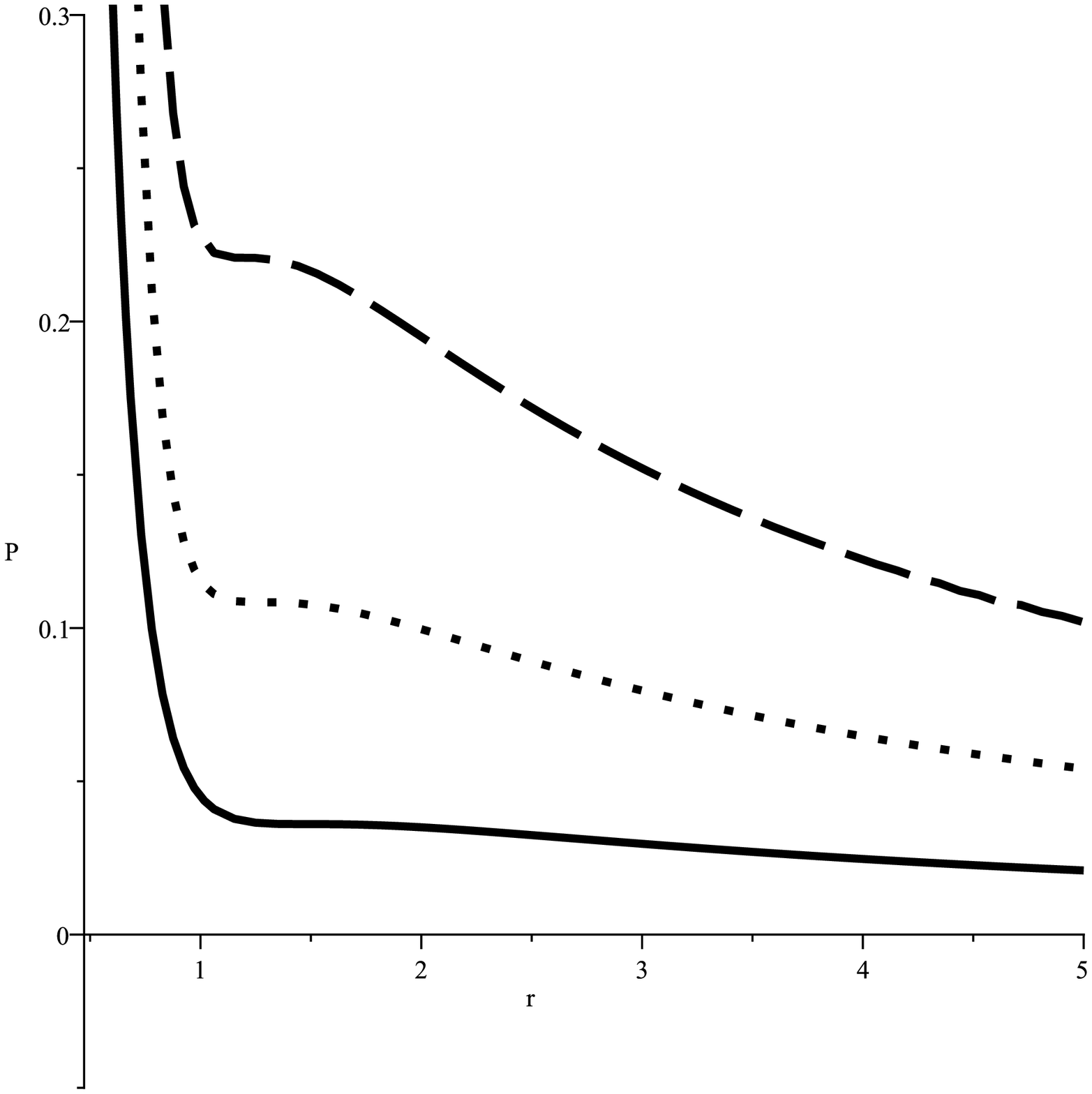} & \epsfxsize=5cm %
\epsffile{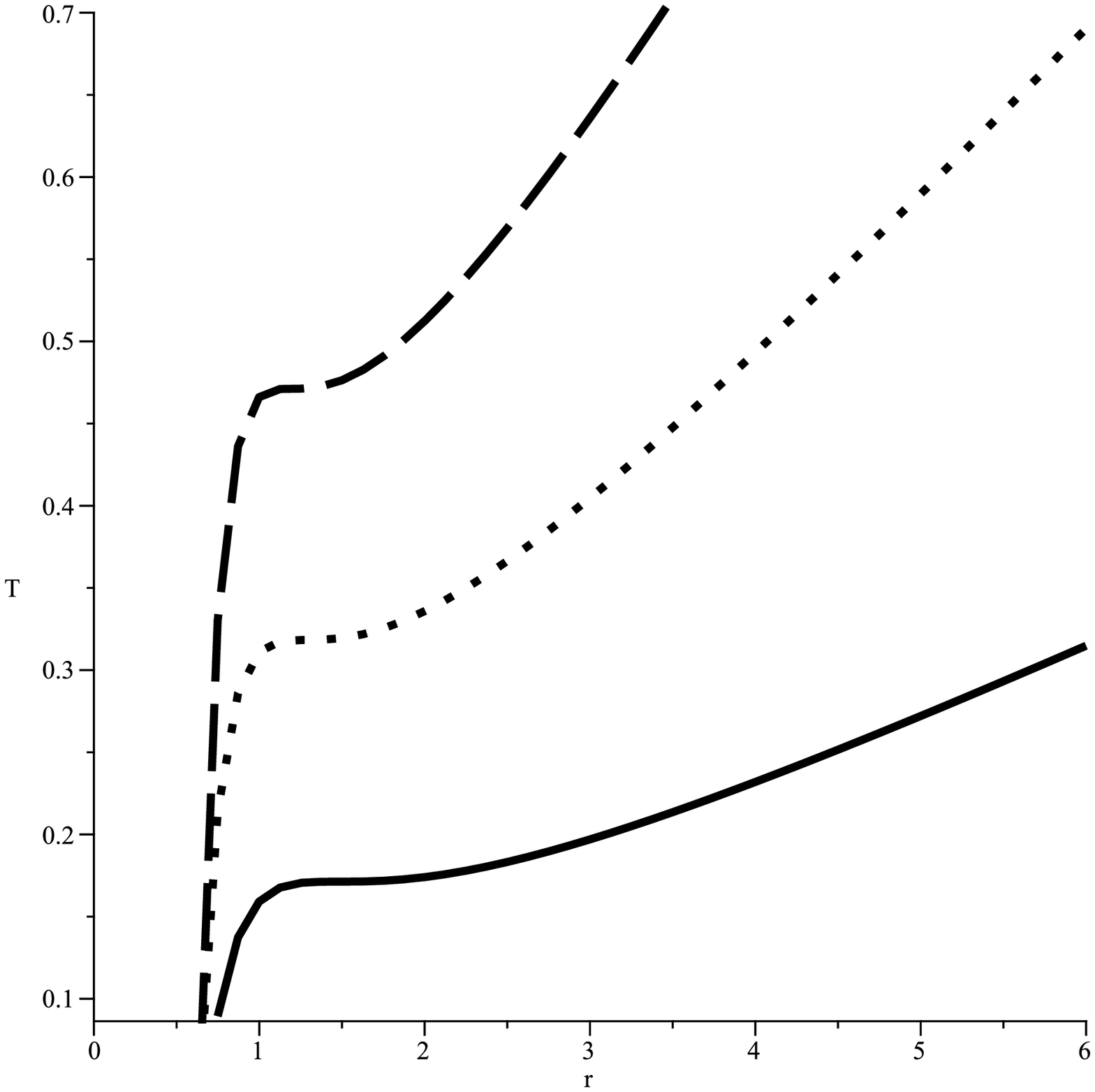} & \epsfxsize=5cm %
\epsffile{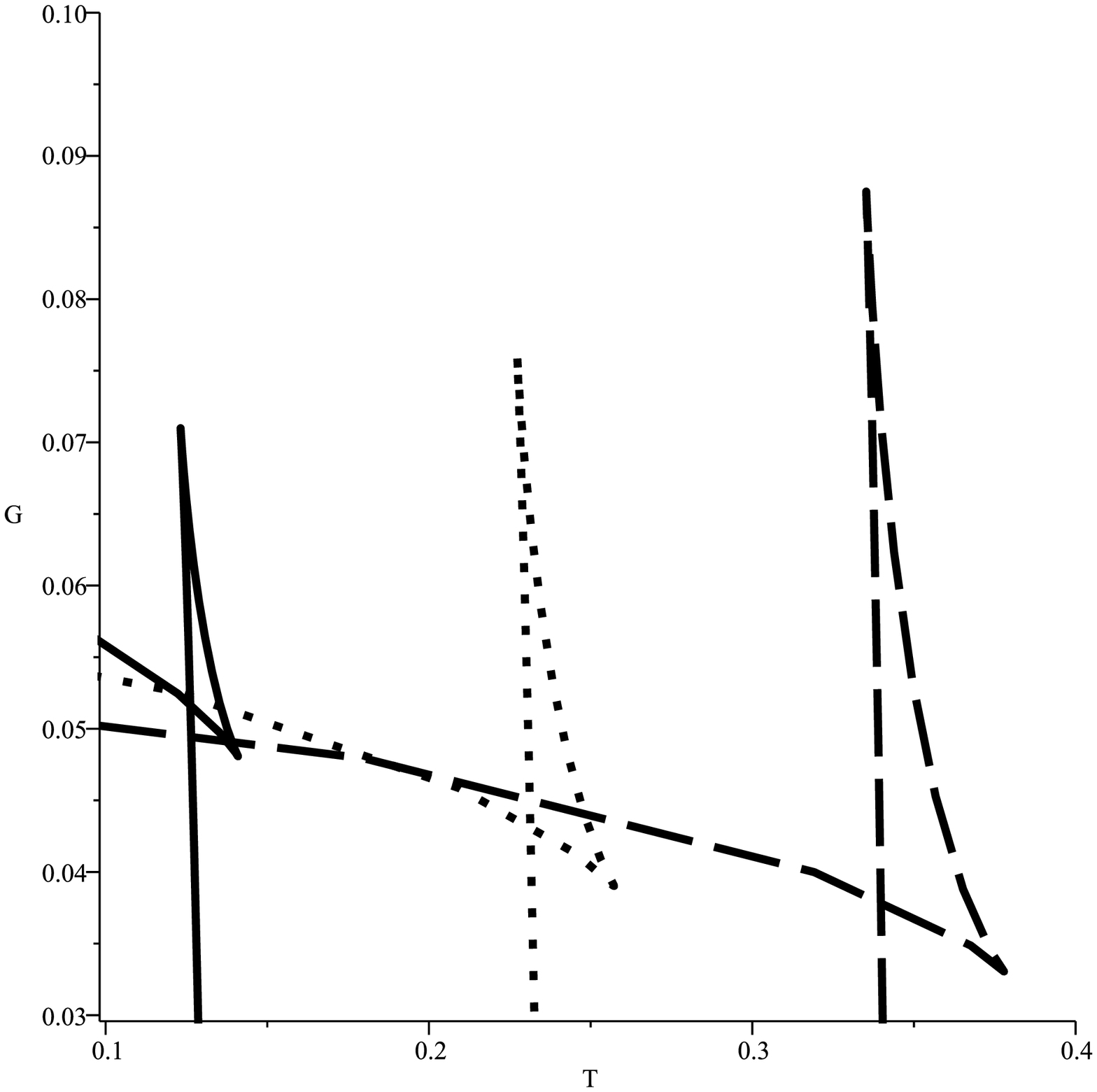}%
\end{array}
$%
\caption{\textbf{"LNEF branch:"} $P-v$ (left), $T-v$ (middle) and $G-T$
(right) diagrams for $q=1$, $P=0.5P_{c}$ and $\protect\beta =1$. \newline
$P-v$, $T-v$ and $G-T$ diagrams for $d=5$ (continuous line), $d=6$ (dot
line) and $d=7$ (dashed line), respectively. }
\label{Fig9Ein}
\end{figure}

Now, we are in a position to analyze the plot of $P-v$ isotherm diagram and
investigate the existence of phase transition. In general, LNEF and ENEF
models for large value of nonlinearity parameter have similar asymptotical
behavior. Moreover, thermodynamical behavior of black hole solutions with
the mentioned models are the same, globally. Therefore, for economical
reason, we will plot phase diagrams for LNEF case. Left diagrams of Figs. %
\ref{Fig2Ein} - \ref{Fig9Ein} indicates an analogue behavior between our
plots with those of Van der Waals gas. One may use the inflection point
properties of critical point to obtain
\begin{eqnarray}
\left( \frac{\partial P}{\partial v}\right) _{T} &=&0,  \label{cr1} \\
\left( \frac{\partial ^{2}P}{\partial v^{2}}\right) _{T} &=&0.  \label{cr2}
\end{eqnarray}

These equations help us to obtain the critical values for the temperature,
the pressure and the volume. Although we cannot, analytically, obtain the
critical values of the temperature, volume and pressures, we may investigate
them numerically. Taking into account the critical quantities, one can
obtain the following universal ratio
\begin{equation}
{\rho }_{c}=\frac{P_{c}v_{c}}{T_{c}}.
\end{equation}

Besides, one may look for the phase transition with the help of $T-v$
diagrams. These graphs are also representing phase transition of black holes
which is of an interest in this paper. Also, with drawing these figures one
can see whether the obtained critical values are essentially critical values
representing phase transition. Another reason for studying $T-v$ diagrams is
due to the fact that comparing to other graphs ($P-v$ and $G-T$), that are
used for studying critical behavior of the system, $T-v$ plots give us
better insight in understanding single phase regions and the effects of
different parameters on these single phase regions. In our case, these
single phase regions are representing small/large (unstable/stable) black
holes. On the other hand, if one is interested in superconductivity that
these nonlinear theories represent, studying of these diagrams gives better
description regarding conductor/superconductor regions and what modifies
them. We plot the mentioned figures in the middle diagram of Figs. \ref%
{Fig2Ein} - \ref{Fig9Ein}. These plots indicate an inflection point which
shows the phase transition of the system. Moreover, phase transition of a
thermodynamical system can be studied by Gibbs free energy. We follow the
method of the extended phase space to obtain Gibbs free energy. The behavior
of the Gibbs free energy with respect to the temperature is displayed in
right diagrams of Figs. \ref{Fig2Ein} - \ref{Fig9Ein}. In these figures the
characteristic swallow-tail behavior guarantees the existence of the phase
transition.


\begin{center}
\begin{tabular}{c}
\begin{tabular}{ccccc}
& Table (1): & $q=1$, & $d=5$. &  \\ \hline\hline
$\beta $ & $v_{c}$ & $T_{c}$ & $P_{c}$ & $\frac{P_{c}v_{c}}{T_{c}}$ \\
\hline\hline
$0.1000$ & $0.4012$ & $0.4004$ & $0.2449$ & $0.2454$ \\ \hline
$0.5000$ & $1.4142$ & $0.1744$ & $0.0375$ & $0.3044$ \\ \hline
$1.0000$ & $1.4765$ & $0.1712$ & $0.0360$ & $0.3106$ \\ \hline
$1.5000$ & $1.4870$ & $0.1707$ & $0.0357$ & $0.3117$ \\ \hline
$2.0000$ & $1.4907$ & $0.1705$ & $0.0356$ & $0.3120$ \\ \hline
\end{tabular}
\\[0pt]
\vspace{0.5cm}%
\end{tabular}

\begin{tabular}{c}
\begin{tabular}{ccccc}
& Table (2): & $q=1$, & $d=7$. &  \\ \hline\hline
$\beta $ & $v_{c}$ & $T_{c}$ & $P_{c}$ & $\frac{P_{c}v_{c}}{T_{c}}$ \\
\hline\hline
$0.1000$ & $0.7380$ & $0.6526$ & $0.4410$ & $0.4987$ \\ \hline
$0.5000$ & $1.1549$ & $0.4769$ & $0.2264$ & $0.5483$ \\ \hline
$1.0000$ & $1.1944$ & $0.4711$ & $0.2207$ & $0.5597$ \\ \hline
$1.5000$ & $1.2014$ & $0.4698$ & $0.2195$ & $0.5612$ \\ \hline
$2.0000$ & $1.2037$ & $0.4694$ & $0.2191$ & $0.5618$ \\ \hline
\end{tabular}
\\[0pt]
.%
\end{tabular}
\end{center}

\section{Discussion on the results of diagrams}

As one can see, thermodynamical behavior of our systems is presented in
Figs. \ref{Fig2Ein} - \ref{Fig9Ein}. The obtained critical values are
representing a phase transition point which is evident from swallowtail
appearing in $G-T$ diagrams. On the other hand, studying $P-v$ and $T-v$
graphs for related critical values shows that obtained values are critical
point that phase transition occurs in them. As nonlinearity parameter
increases, the temperature, in which phase transitions take place, decreases
and the value of Gibbs free energy of phase transition points increases.
These results indicate that in more powerful nonlinearity parameter our
thermodynamical system needs less energy in order to have a phase transition
(see the middle and right diagrams of Fig. \ref{Fig8Ein} for more details).
This could also be interpreted from the fact that as nonlinearity parameter
increases, the Enthalpy which is represented by total finite mass of black
hole increases, too. Therefore, for obtained black holes we expect to phase
transition occurs with absorbing less mass form surrounding.

Studying $P-v$ diagram shows that as $\beta $ increases, the critical
pressure decreases, but in opposite side the horizon radius of critical
points increases. Also, one can see that as the nonlinearity parameter
increases the distance between two diagrams related to two different values
of nonlinearity decreases which shows the fact that the effect of
nonlinearity in higher values do not change critical values so much (see the
left diagrams of Figs. \ref{Fig7Ein} and \ref{Fig8Ein} for more details).
Also, one can argue that due to fact that pressure is related to
cosmological constant which is related to asymptotical curvature of the
background and our thermodynamical system, as nonlinearity parameter
increases, the necessity of having a background with higher curvature
decreases.

As one can see in $T-v$ diagrams, as the nonlinearity parameter increases,
temperature (horizon radius) of critical points of phase transition
decreases (increases) (see the middle diagram of Figs. \ref{Fig7Ein} and \ref%
{Fig8Ein}). In other words, in order to have phase transition for large $%
\beta$, we need less energy which is consistent with results that we find in
studying Gibbs free energy diagrams. Therefore, for having phase transition
in the presence of higher value of $\beta$, black hole needs to absorb less
mass in order to achieve stable state.

The effects of dimensionality on critical points and their behavior is
another interesting issue that we discuss in this section. The obtained $G-T$
diagram for different dimensions shows that as dimensionality increases, the
temperature of critical points increases which indicates the necessity of
more energy for having a phase transition (see the right diagram of Fig. \ref%
{Fig9Ein} for more details). On the other hand, the gap between two
different phases of our thermodynamical system increases which shows the
fact that as dimensionality increases the change in energy of system in
which phase transition occurs, increases too. For the $P-v$ diagrams, we
have higher pressure in which phase transition takes place. In other words,
as dimensions of system increase pressure, hence cosmological constant
increases, which is acceptable because of the fact that cosmological
constant is dimension dependent (see the left diagram of Fig. \ref{Fig9Ein}
for more details). Finally, for the $T-v$ diagram of Fig. \ref{Fig9Ein}, we
can see that the temperature of critical value and the length of subcritical
isobars increases which means that the single phase region of small/large
black holes decreases. Therefore, as dimensionality increases the system
needs to absorb more mass in order to have phase transition.

Finally, we are studying the behavior of critical values and the universal
ratio of $\frac{P_{c}v_{c}}{T_{c}}.$ As One can see, the critical horizon
radius increases, as nonlinearity parameter increases, whereas critical
temperature and pressure decrease. Also, we can see the ratio of $\frac{%
P_{c}v_{c}}{T_{c}}$ is an increasing function of nonlinearity parameter. For
higher dimensions, we have higher values of critical points which result
into higher value of $\frac{P_{c}v_{c}}{T_{c}}$. Studying tables also
confirms the results that we have derived through studying graphs. As the
nonlinearity parameter increases, the thermodynamical system will be in need
of less energy (in our black hole case by absorbing mass) in order to have
phase transition.

\section{Heat Capacity and GTD in Extended phase space}

In this section, we expand our study regarding the critical behavior of the
system in context of heat capacity and GTD. In order to study the critical
behavior of the system, there are several approaches. One of these
approaches is studying the behavior of the heat capacity. It is arguable
that there are two types of phase transition: one is related to changes in
the signature of the heat capacity. In other words, roots of the heat
capacity are representing phase transitions which we will call them type
one. The other type of phase transition which we will call it type two, is
related to divergency of the heat capacity. It means that singularities of
the heat capacity are representing places in which system goes under phase
transition. For the heat capacity in extended phase space we have following
relation
\begin{equation}
C_{Q,P}=T\left( \frac{\partial S}{\partial T} \right) _{Q,P}.  \label{HC}
\end{equation}

On the other hand, another approach for studying the thermodynamical
behavior of the system is through thermodynamical metric. In other words,
one can use a thermodynamical potential with specific set of extensive
parameters in order to construct a metric. The Ricci scalar of this metric
may contain divergence points in which phase transitions take place in them.
It was shown that in order to this approach and heat capacity approach lead
into same results, the divergencies of thermodynamical Ricci scalar (TRS)
and phase transitions of the heat capacity (regardless of their type) must
coincide.

The problem in constructing an effective GTD model for studying phase
transitions in extended phase is the fact that $M(S,Q,P)$ is a linear
function with respect to the pressure ($P$) whereas different GTD methods
contain second derivation with respect to extensive parameters. Therefore,
using these usual methods result into Ricci scalar of the constructed metric
being zero (since $M_{PP}=\frac{\partial ^{2}M}{\partial P^{2}}=0$). To
summarize, we need a model in which:

\emph{\textbf{1.} All the divergence points of Ricci scalar of constructed
spacetime, coincide with phase transition points of the heat capacity. }

\emph{\textbf{2.} It must contain at last up to first order derivation with
respect to extensive parameters (such as pressure).}

In order to making considered GTD model more effective, we will add third
condition for the GTD model:

\emph{\textbf{3.} The critical behavior of the Ricci scalar of the
constructed spacetime must be consisting with our critical behavior that is
observed in extended phase space. In other words, characteristic behavior
that is observed for critical values in studying $P-v$, $T-v$ and $G-T$ must
be seen in the behavior of Ricci scalar. The critical behavior in case of $%
P<P_{c}$, $P=P_{c}$ and $P>P_{c}$ must be consistent in all these three
approaches.}

Armed with these three conditions, we will introduce the following metric
for our case of study (Case $I$)
\begin{equation}
ds^{2}=S\frac{M_{S}}{M_{QQ}^{3}}\left(
-M_{SS}dS^{2}+M_{QQ}dQ^{2}+dP^{2}\right) ,  \label{New}
\end{equation}
where $M_{J}=\frac{\partial M}{\partial J}$ and $M_{JJ}=\frac{\partial ^{2}M%
}{\partial J^{2}}$. For economical reasons we will not bring obtained values
for the heat capacity, Eq. (\ref{HC}), and also TRS of the constructed
thermodynamical metric, Eq. (\ref{New}). Since we are looking for the
divergence points of TRS, it is sufficient to investigate its denominator.
It is a matter of calculation to show that the denominator of TRS has the
following form
\begin{equation}
denom(\mathcal{R})=S^{3}M_{S}^{3}M_{SS}^{2}.
\end{equation}

\begin{figure}[tbp]
$%
\begin{array}{ccc}
\epsfxsize=5cm \epsffile{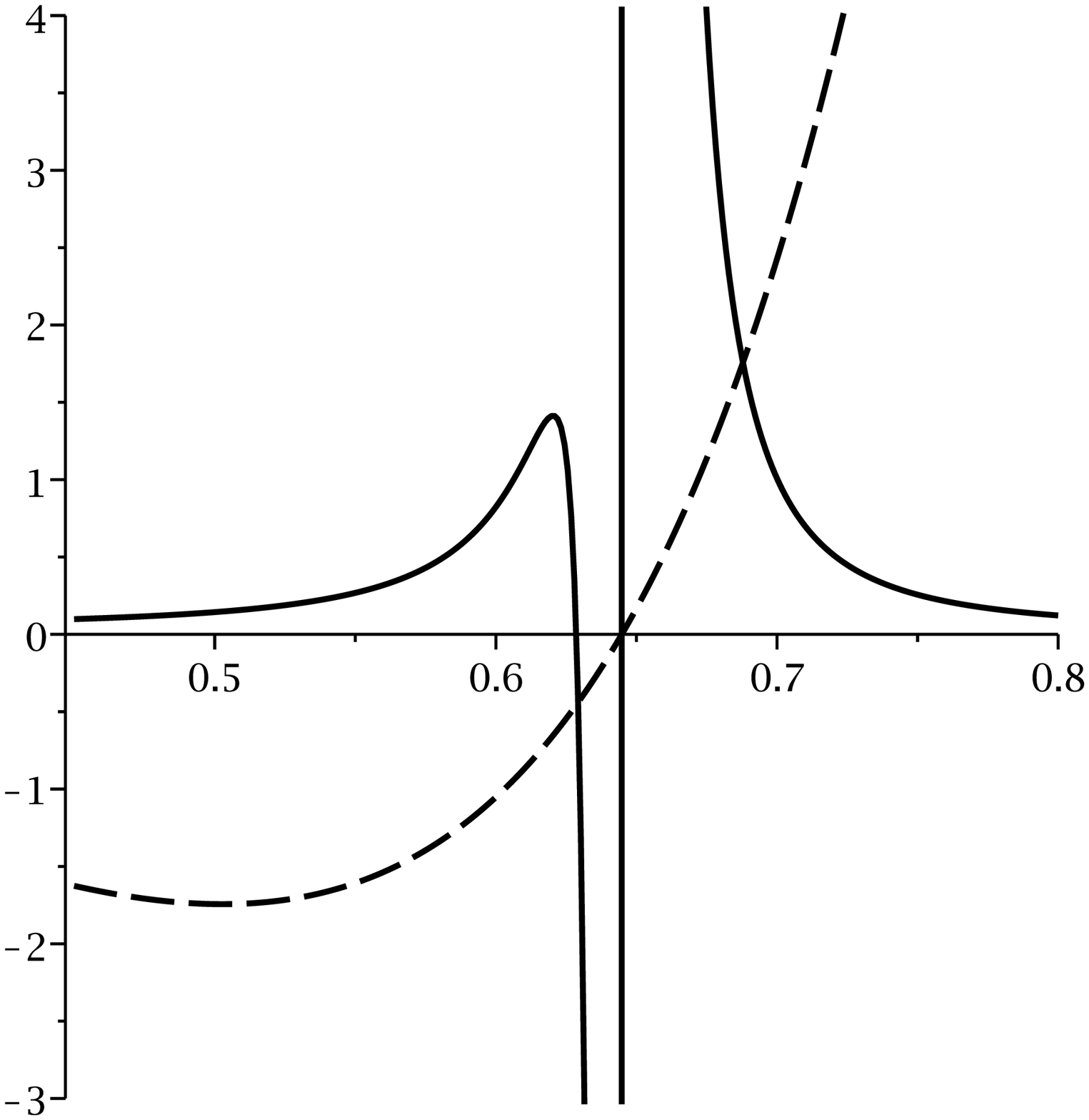} & \epsfxsize=5cm %
\epsffile{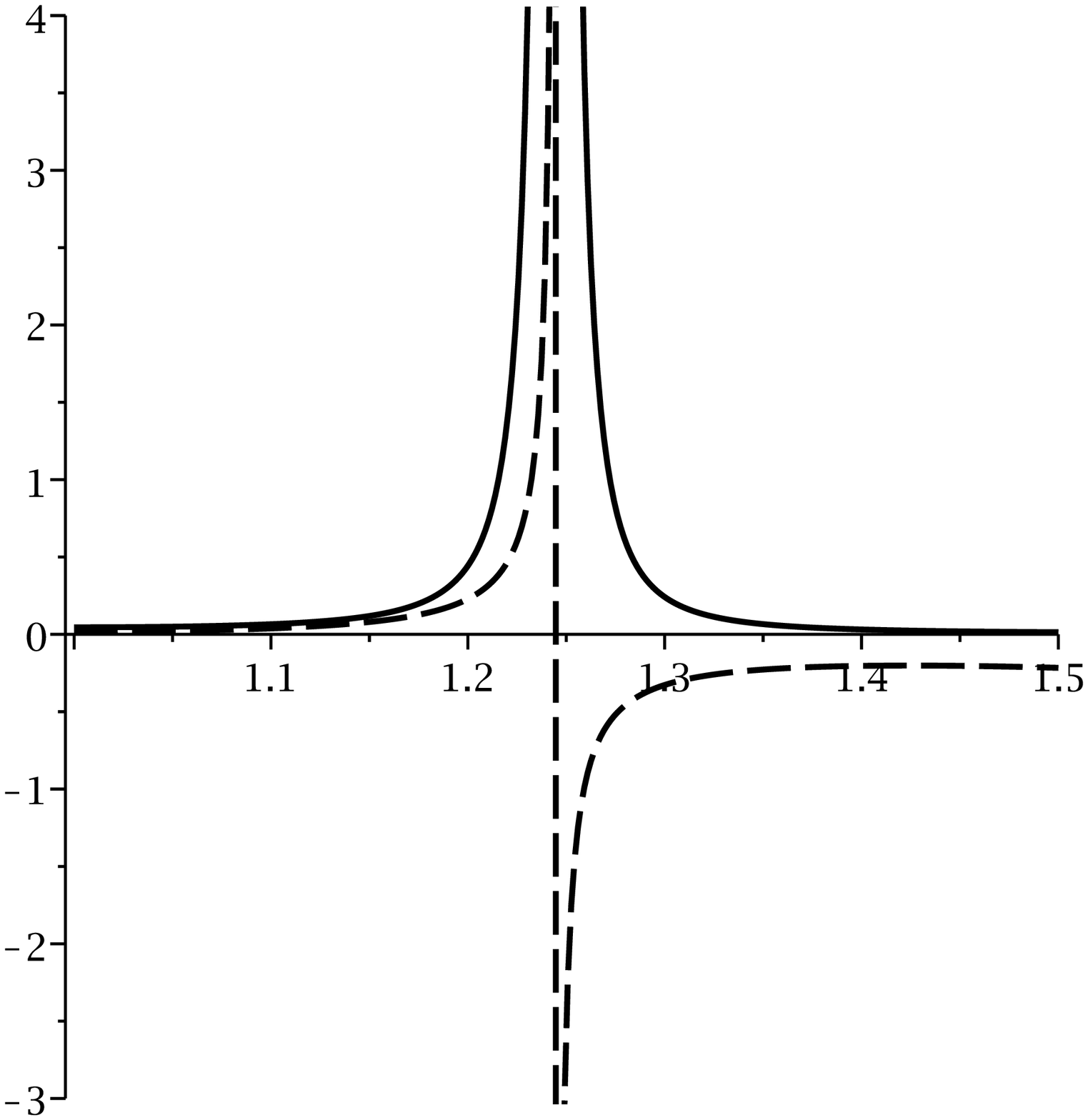} & \epsfxsize=5cm \epsffile{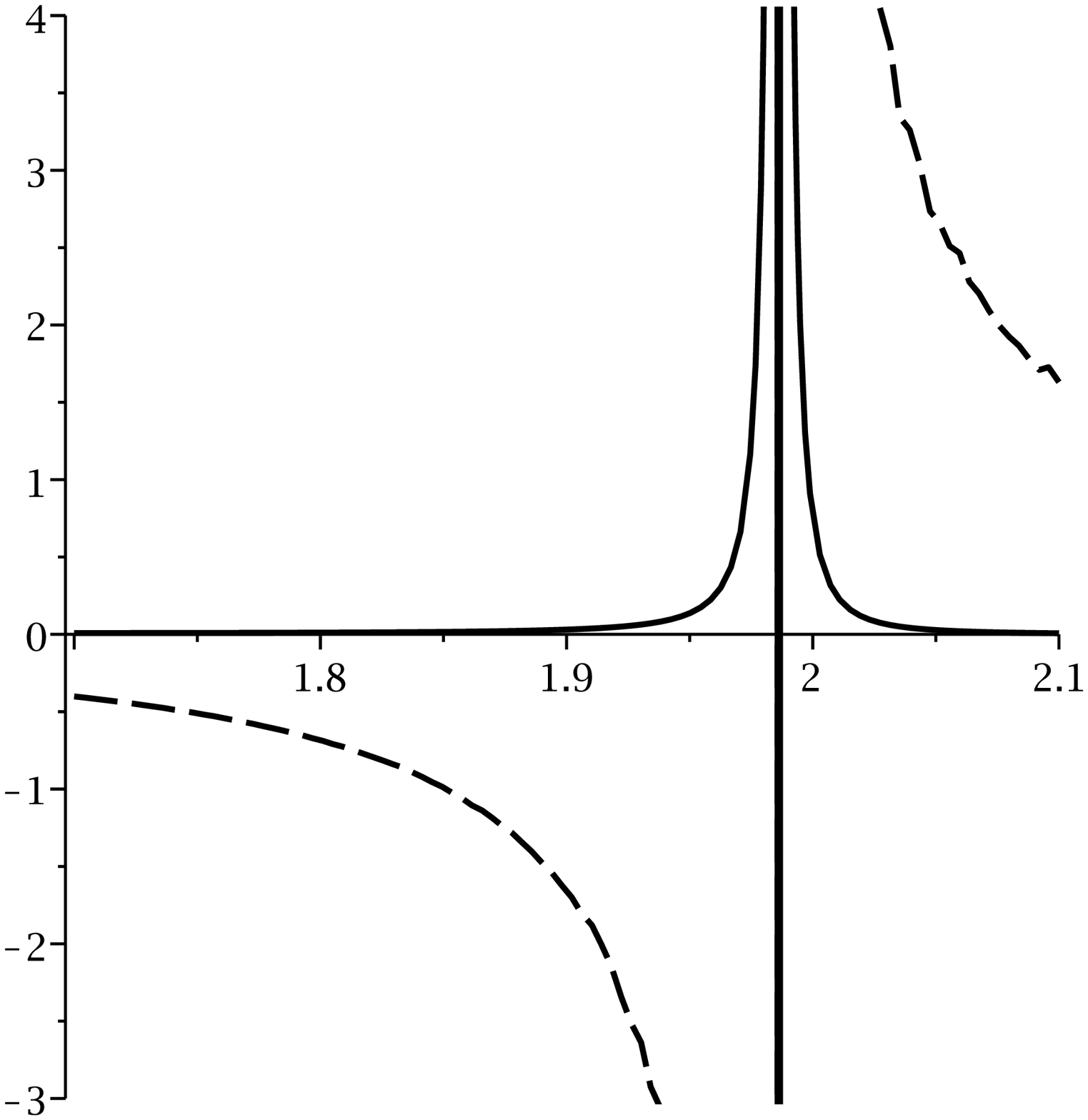}%
\end{array}
$%
\caption{\textbf{"LNEF branch:"} $\mathcal{R}$ (continuous line - Case $I$),
$C_{Q}$ (dashed line) diagrams for $d=5$, $q=1$ and $\protect\beta=1$.
\newline
for different scales: $P=0.75P_{c}$. \newline
.}
\label{101}
\end{figure}

\begin{figure}[tbp]
$%
\begin{array}{ccc}
\epsfxsize=5cm \epsffile{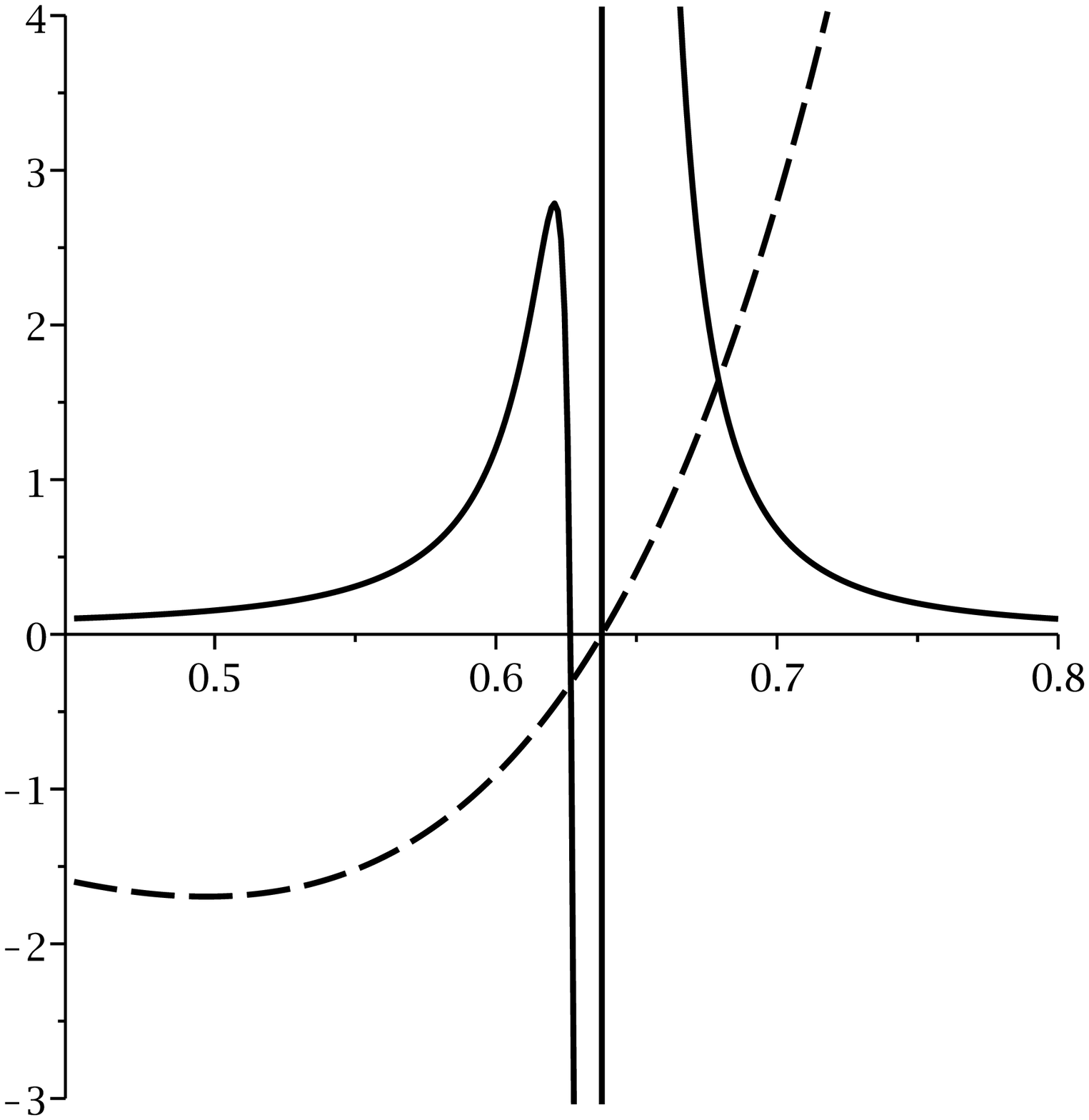} & \epsfxsize=5cm \epsffile{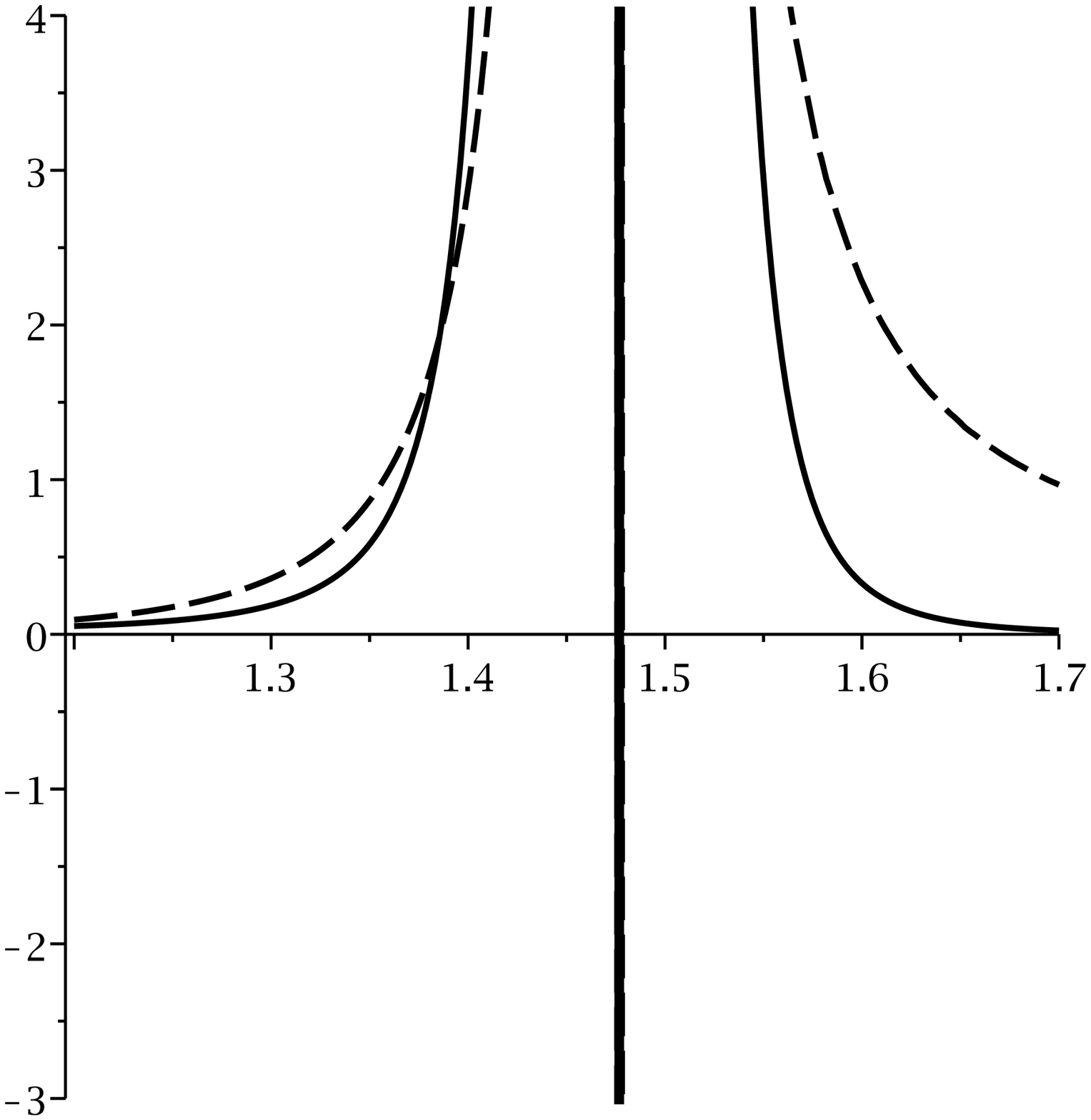}
& \epsfxsize=5cm \epsffile{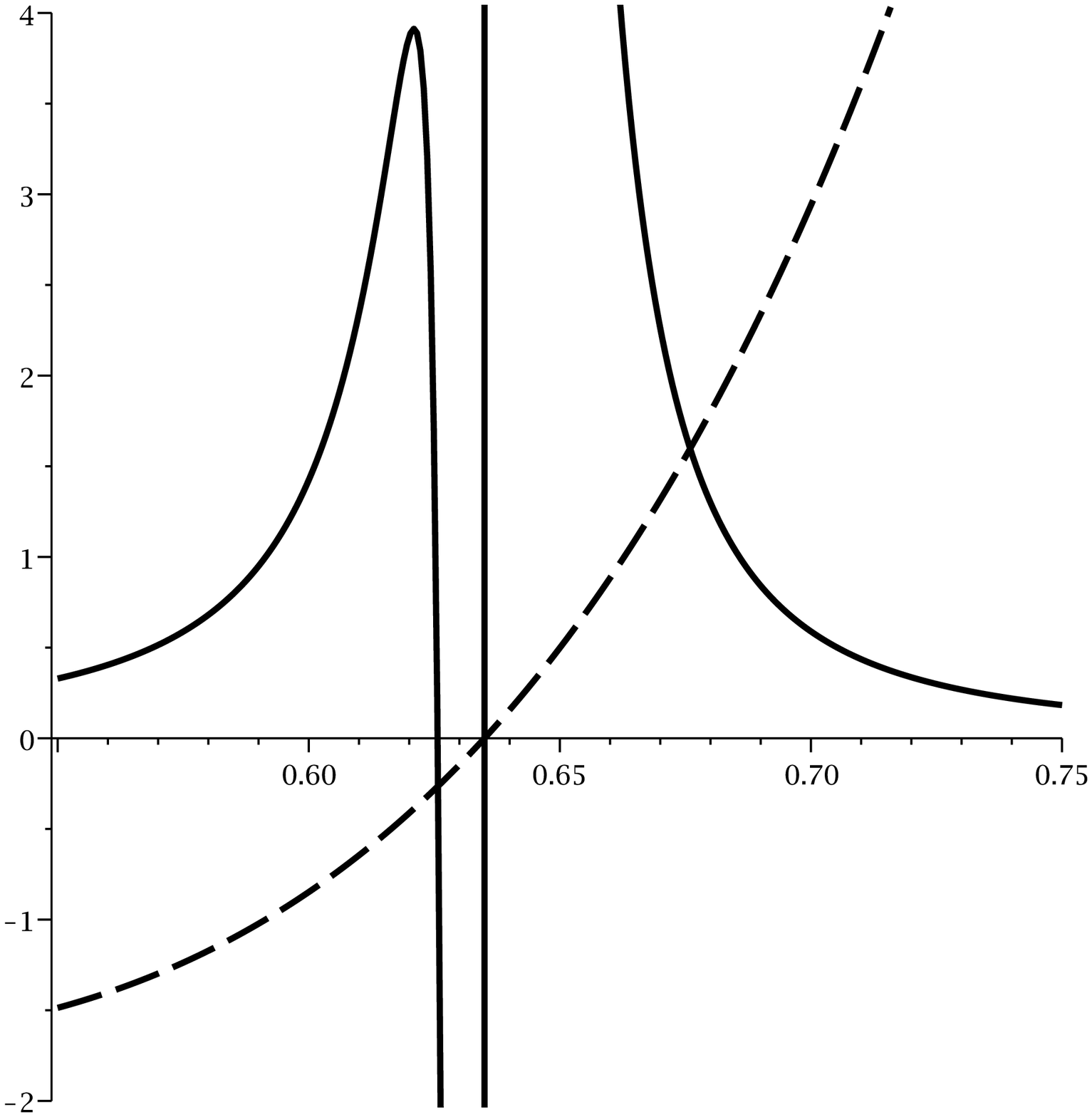}%
\end{array}
$%
\caption{\textbf{"LNEF branch:"} $\mathcal{R}$ (continuous line - Case $I$),
$C_{Q}$ (dashed line) diagrams for $d=5$, $q=1$ and $\protect\beta=1$.
\newline
for $P=P_{c}$ middle and left (different scales) and $P=1.1P_{c}$ Right.
\newline
.}
\label{102}
\end{figure}

\begin{figure}[tbp]
$%
\begin{array}{ccc}
\epsfxsize=5cm \epsffile{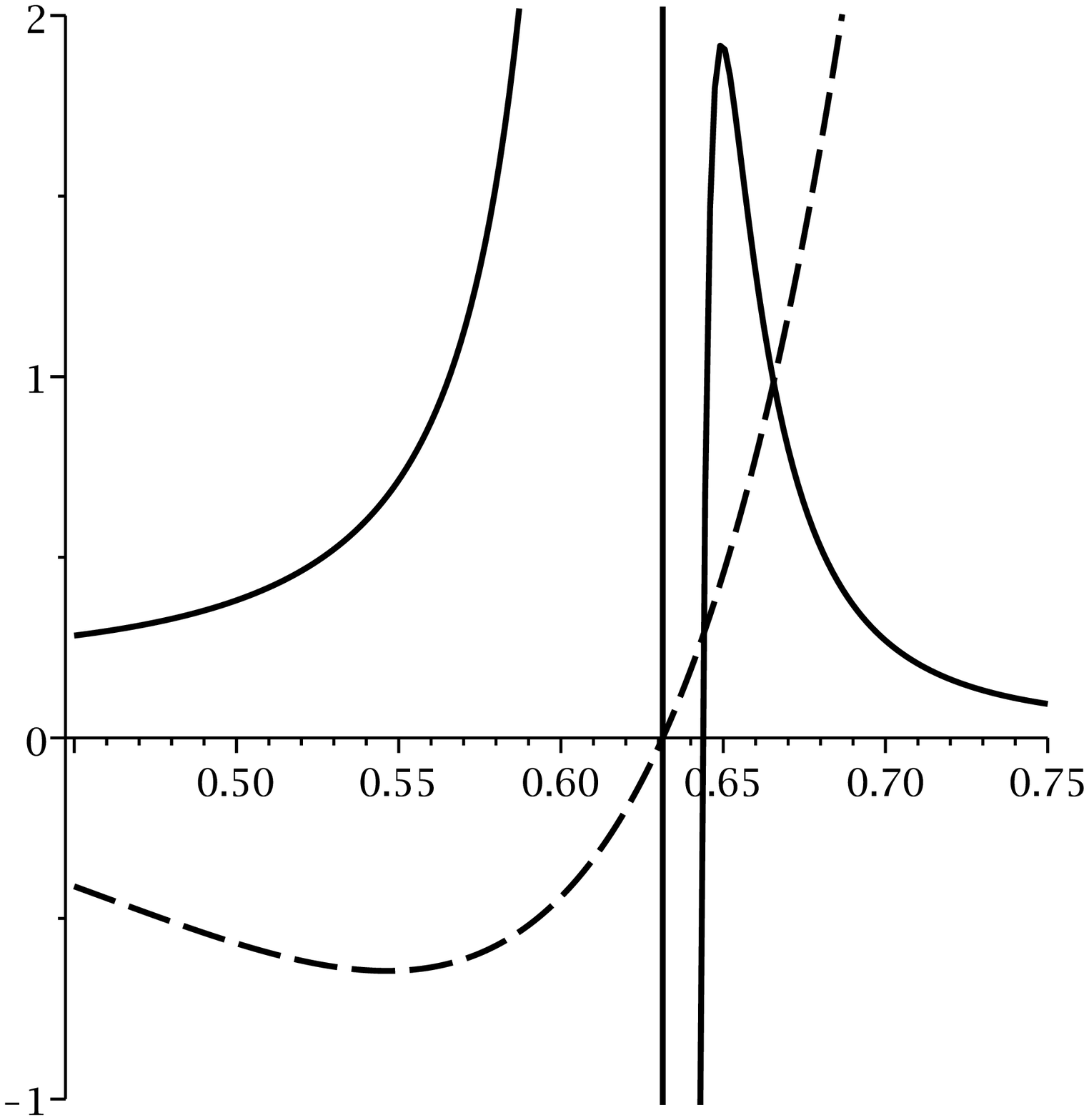} & \epsfxsize=5cm %
\epsffile{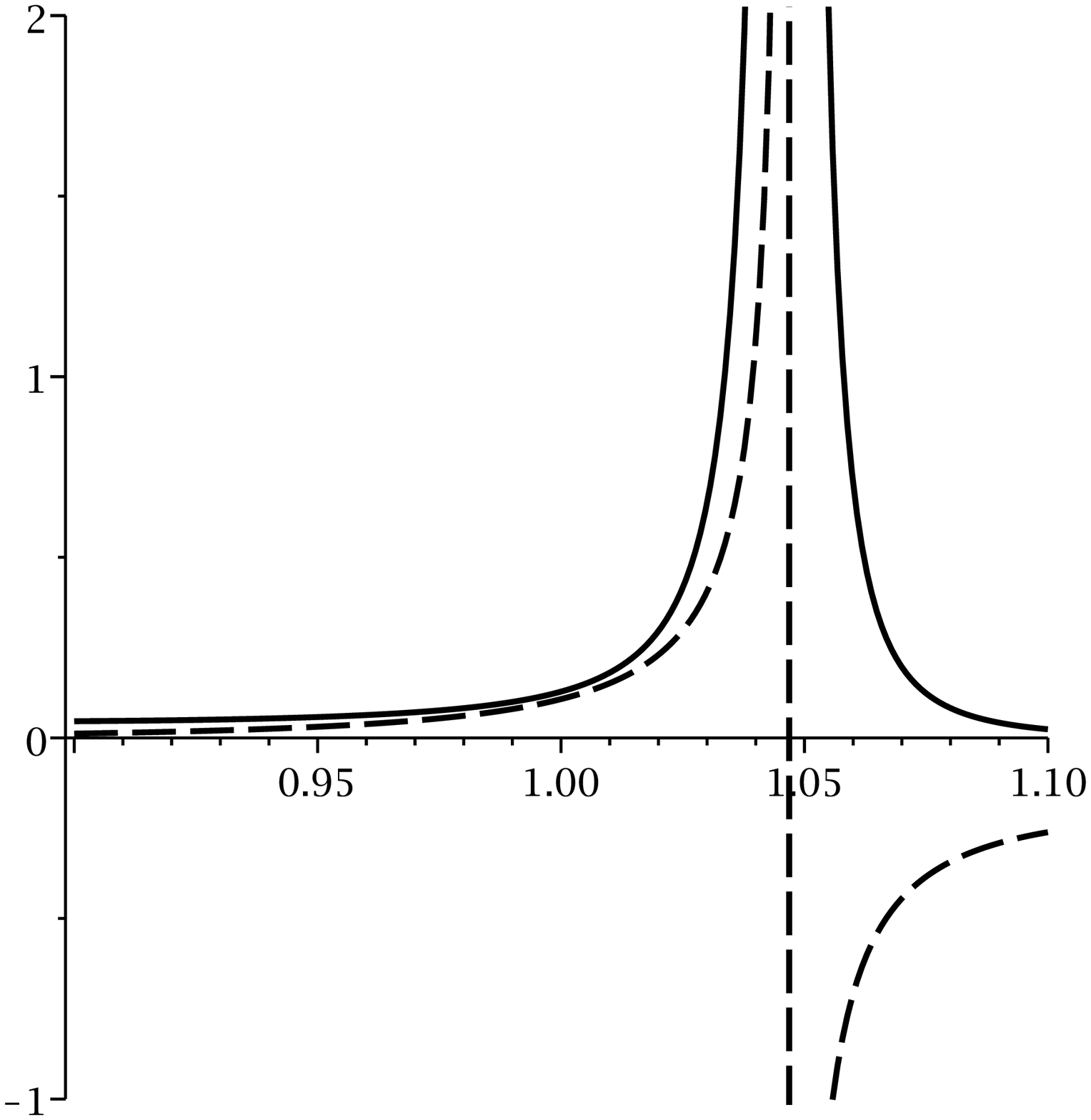} & \epsfxsize=5cm \epsffile{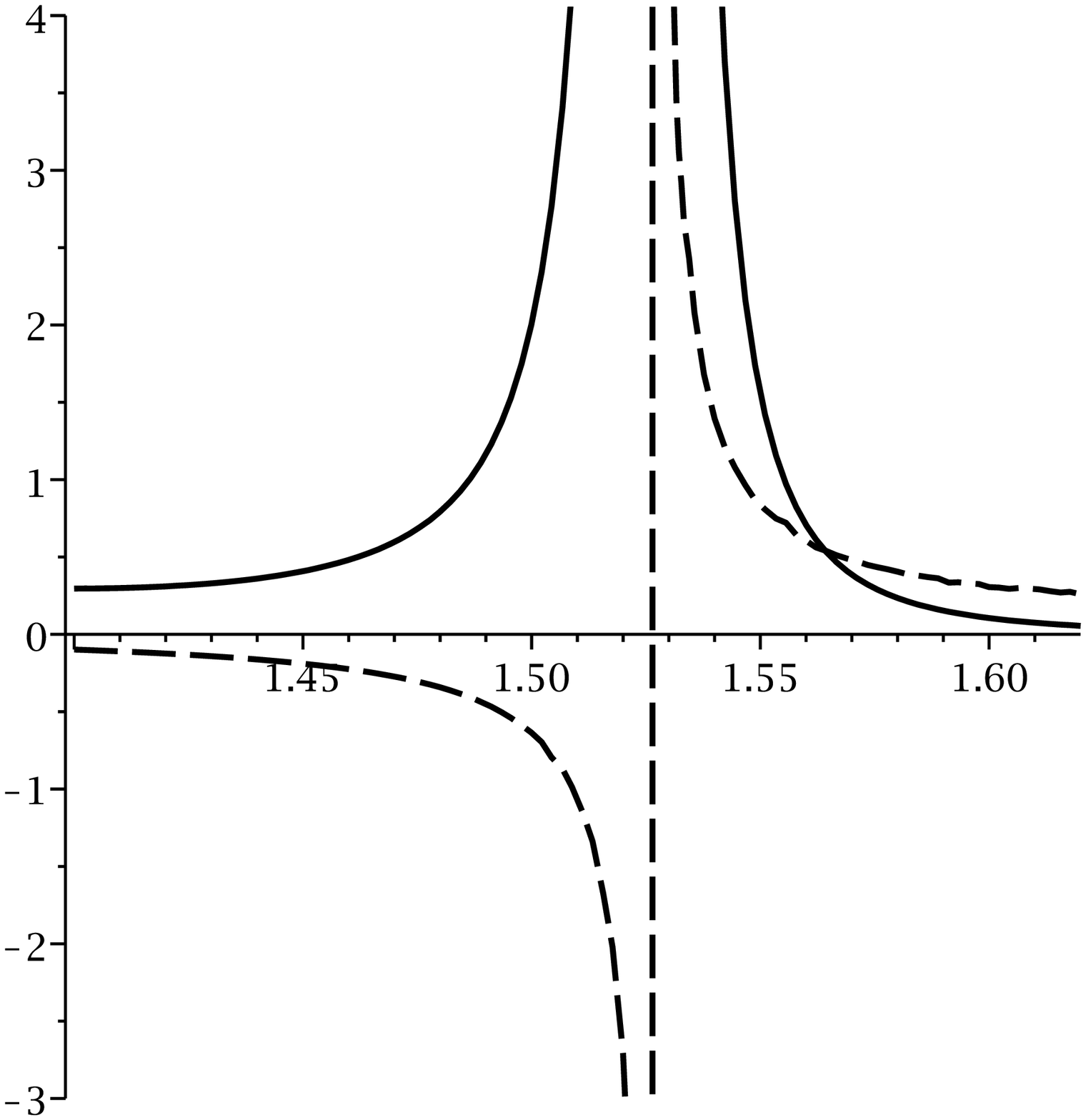}
\end{array}
$%
\caption{\textbf{"LNEF branch:"} $\mathcal{R}$ (continuous line - Case $I$),
$C_{Q}$ (dashed line) diagrams for $d=7$, $q=1$ and $\protect\beta=1$.
\newline
for different scales: $P=0.75P_{c}$.}
\label{111}
\end{figure}

\begin{figure}[tbp]
$%
\begin{array}{ccc}
&  &  \\
\epsfxsize=5cm \epsffile{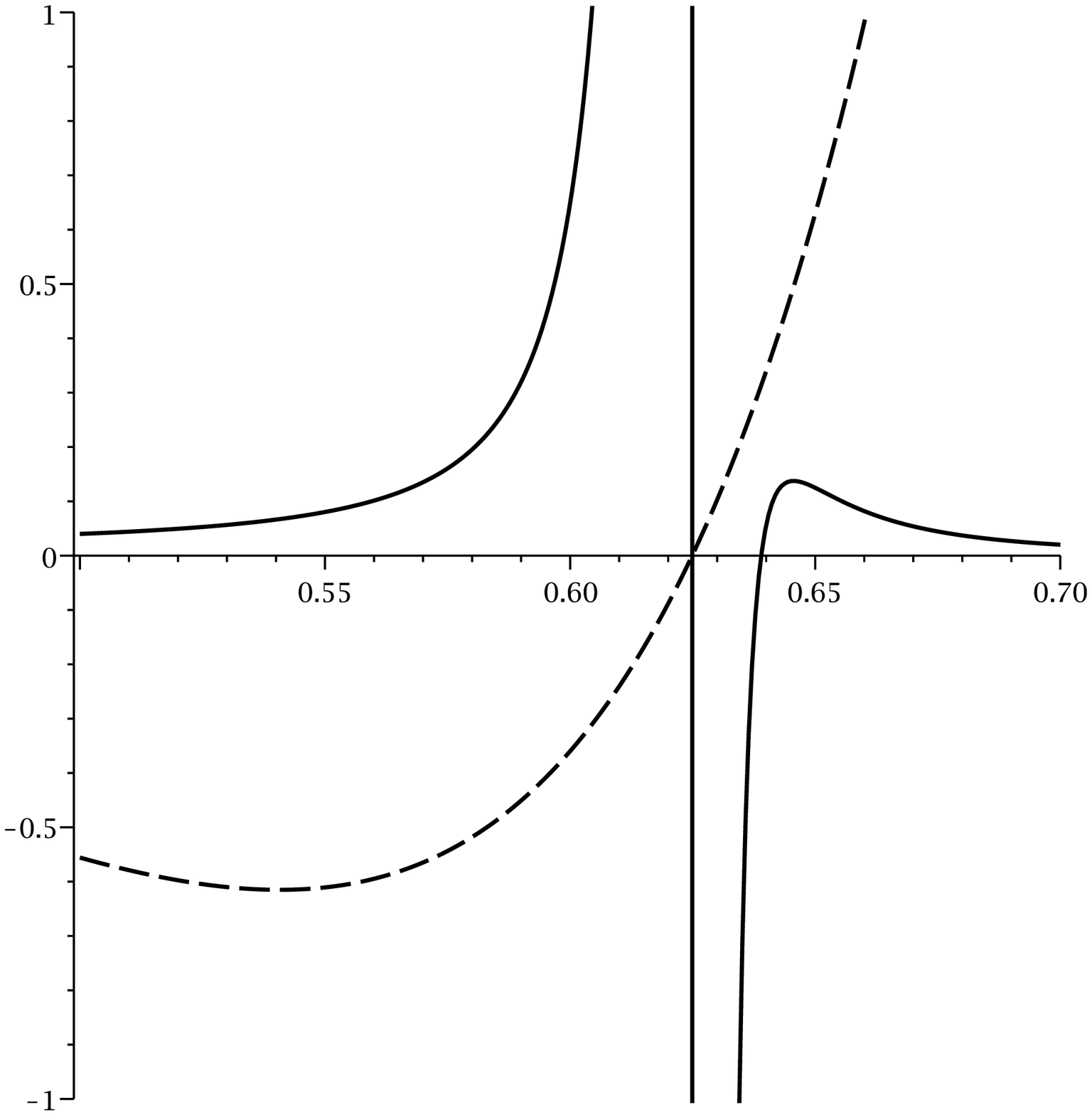} & \epsfxsize=5cm \epsffile{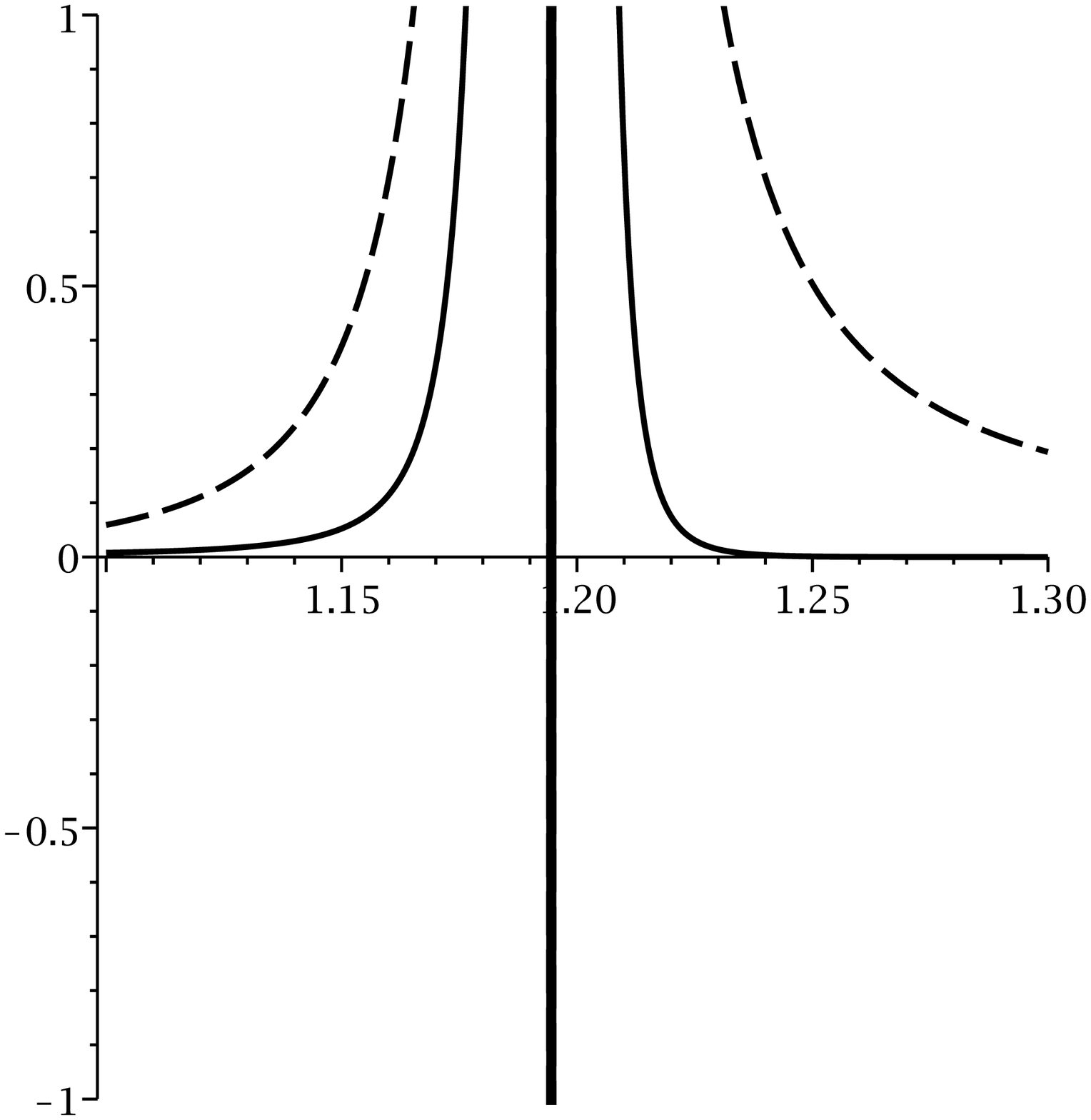}
& \epsfxsize=5cm \epsffile{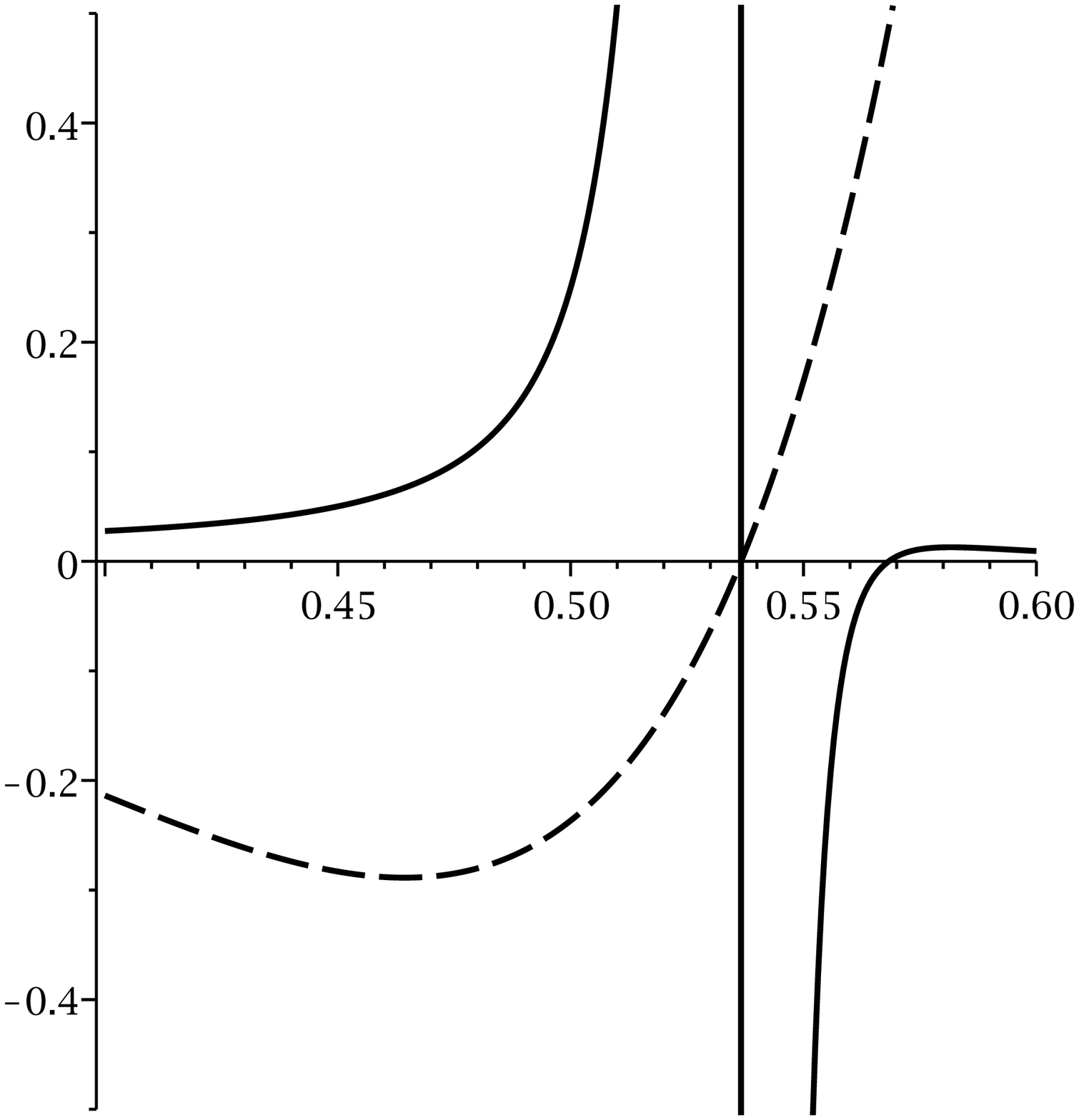}%
\end{array}
$%
\caption{\textbf{"LNEF branch:"} $\mathcal{R}$ (continuous line - Case $I$),
$C_{Q}$ (dotted line) diagrams for $d=7$, $q=1$ and $\protect\beta=1$.
\newline
for $P=P_{c}$ middle and left (different scales) and $P=1.1P_{c}$ Right.}
\label{112}
\end{figure}

As one can see, the denominator of TRS contains both numerator and
denominator of the heat capacity ($M_{S}=T$ and $M_{SS}$). Therefore, our
first condition is satisfied. On the other hand, the constructed metric has
more than first derivative with respect to extensive parameters and its
Ricci scalar is nonzero. Therefore, our second condition is also satisfied.
As for the final condition, in order to elaborate the effectiveness of this
thermodynamical metric, we will plot some diagrams with respect to obtained
critical values in previous sections (Figs. \ref{101} - \ref{112}).

First of all, we should point out that there is one divergency for TRS which
is related to root of the heat capacity in all plotted diagrams. As one can
see, for the case of $P=P_{c}$ (left and middle panels of Figs. \ref{102}
and \ref{112}), there is a divergency for heat capacity which is located
exactly at critical horizon radius of extended phase space (compare it with
table 1 and 2, and middle diagrams of Figs. \ref{Fig2Ein} and \ref{Fig5Ein}%
). Also for this case, both TRS and $C_{Q}$ have only one divergency at the
same place. As for $P<P_{c}$ (Figs. \ref{101} and \ref{111}), there are two
divergencies for the heat capacity and TRS which coincide with each other.
Finally, for the case of $P>P_{c}$, no divergency is observed for the heat
capacity, TRS and $T-v$ diagrams of extended phase space (middle panels of
Figs. \ref{Fig2Ein} and \ref{Fig5Ein}, right panels of Figs. \ref{102} and %
\ref{112}).

As one can see employed thermodynamical metric is providing an effective
machinery in which the behaviors of the heat capacity and phase diagrams of
the extended phase space are seen. In other words, constructed spacetime
posses characteristics and properties in which enable it to present all the
critical behavior of the heat capacity plus extended phase space at the same
time. It is worthwhile to mention that characteristic behavior of TRS
enables one to both types of the phase transitions from each other. For more
clarifications, we should note that for type two phase transitions, the sign
of TRS remains fixed around the divergence points of the heat capacity,
while in the case of type one, the sign of TRS will be changed around the
root of $C_{Q}$. In addition, we should note that the critical points of the
extended phase space are second order phase transitions and they match to
the divergence points of the heat capacity.

Considering mentioned conditions for constructing thermodynamical metric,
one can introduce another type of metric which is similar with the previous
one with one difference. This metric has the following structure (Case $II$)
\begin{equation}
ds^{2}=S\frac{M_{S}}{M_{QQ}^{3}}\left(
-M_{SS}dS^{2}+M_{QQ}dQ^{2}+M_{P}dP^{2}\right) ,  \label{New2}
\end{equation}
in which, its Ricci scalar denominator will be in the following form
\begin{equation}
denom(\mathcal{R})=S^{3}M_{S}^{3}M_{SS}^{2}M_{P}^{2}.
\end{equation}

\begin{figure}[tbp]
$%
\begin{array}{ccc}
\epsfxsize=5cm \epsffile{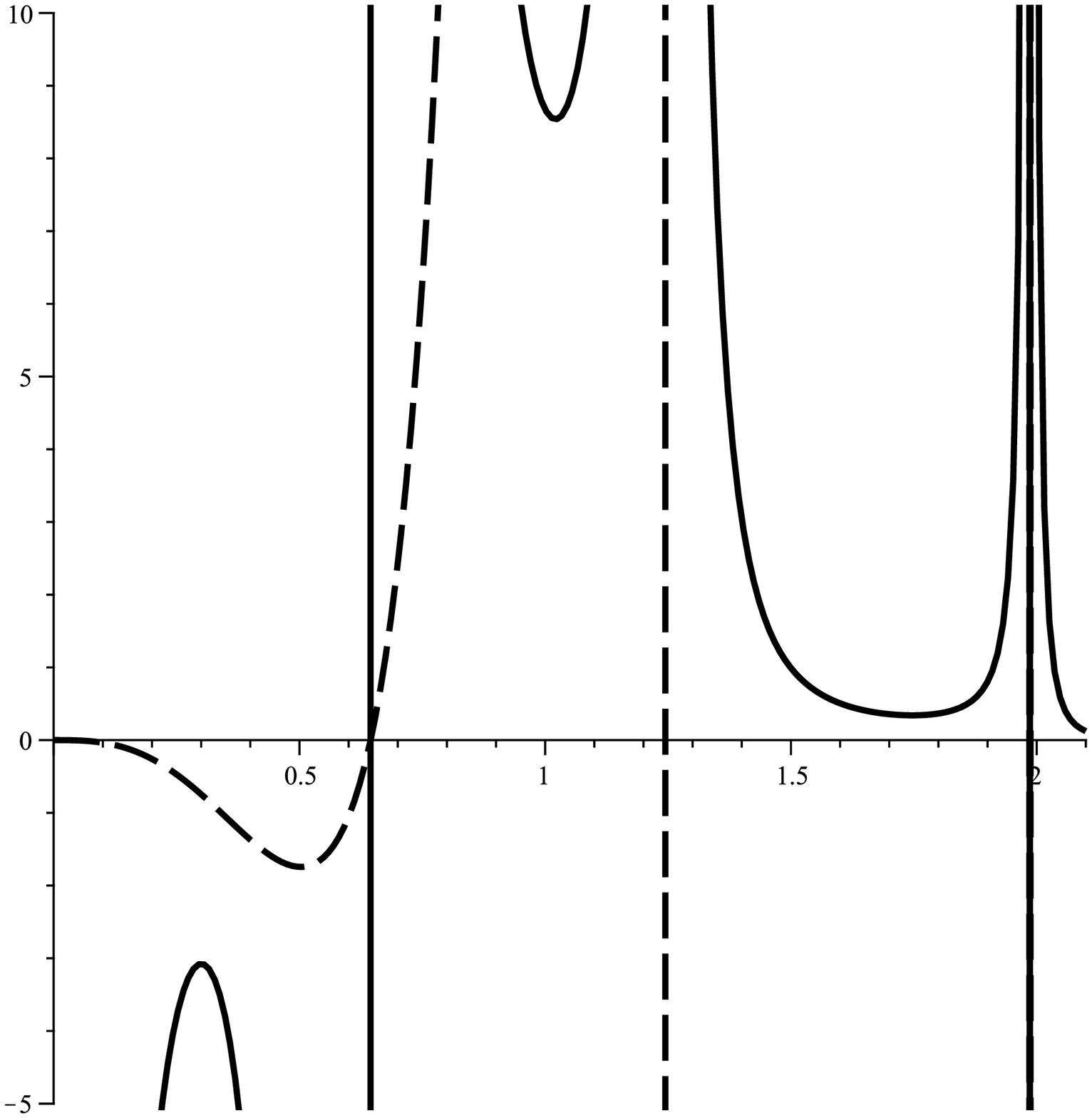} & \epsfxsize=5cm %
\epsffile{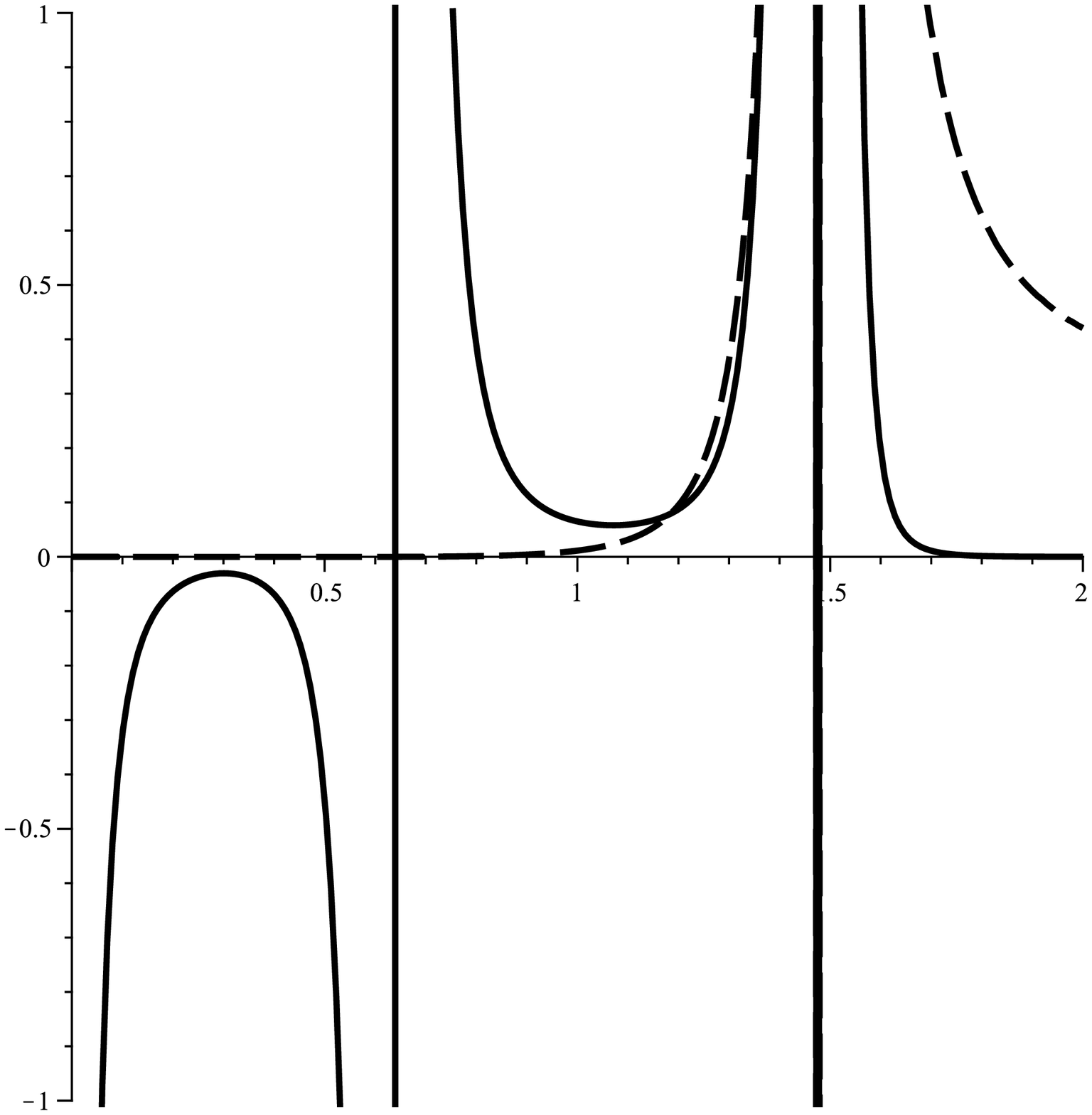} & \epsfxsize=5cm \epsffile{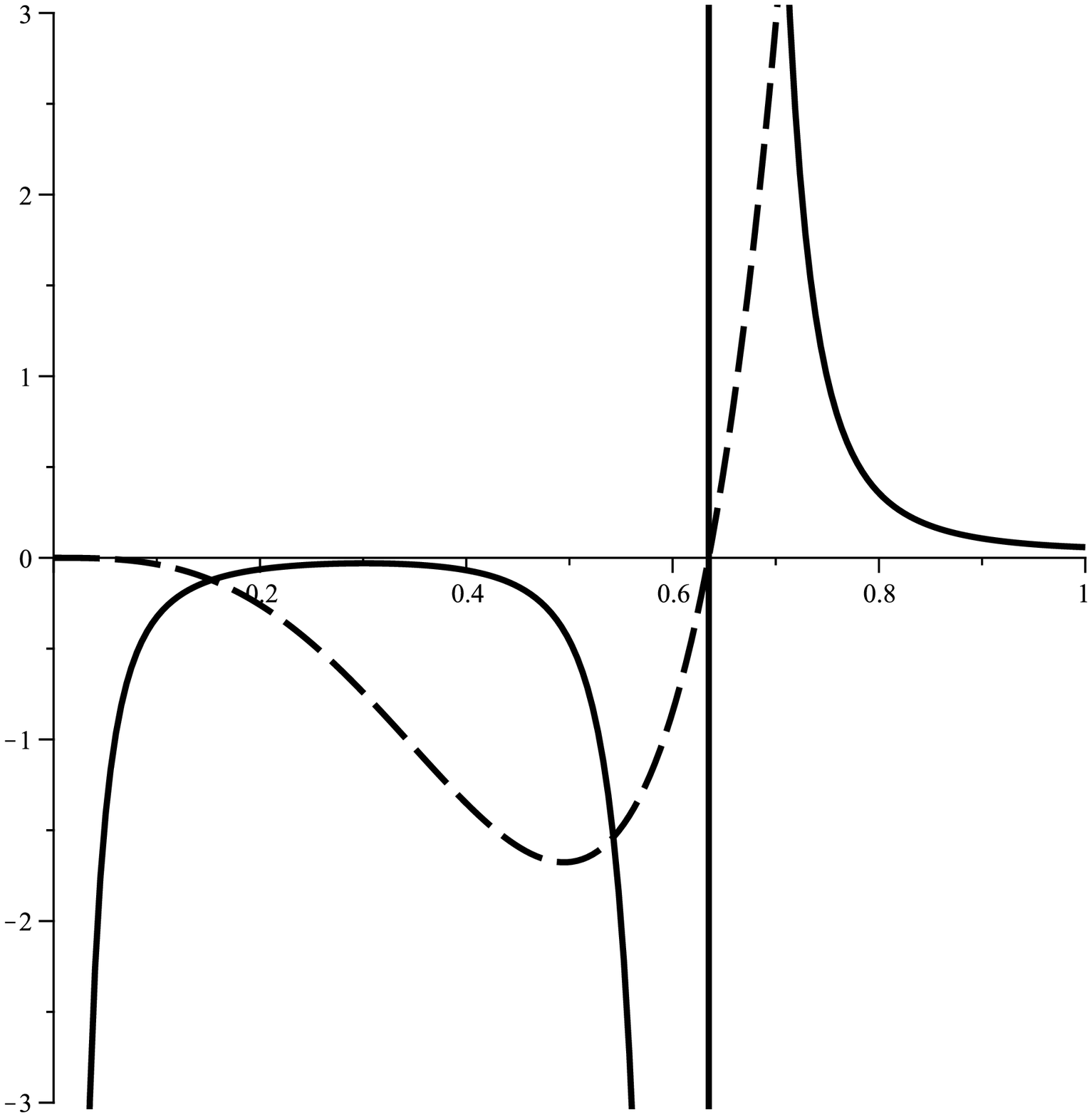}%
\end{array}
$%
\caption{\textbf{"LNEF branch:"} $\mathcal{R}$ (continuous line - Case $II$%
), $C_{Q}$ (dashed line) diagrams for $d=5$, $q=1$ and $\protect\beta =1$.
\newline
for $P=0.75P_{c}$ left, $P=P_{c}$ middle, $P=1.1P_{c}$ Right and . \newline
.}
\label{12}
\end{figure}

\begin{figure}[tbp]
$%
\begin{array}{cc}
\epsfxsize=6cm \epsffile{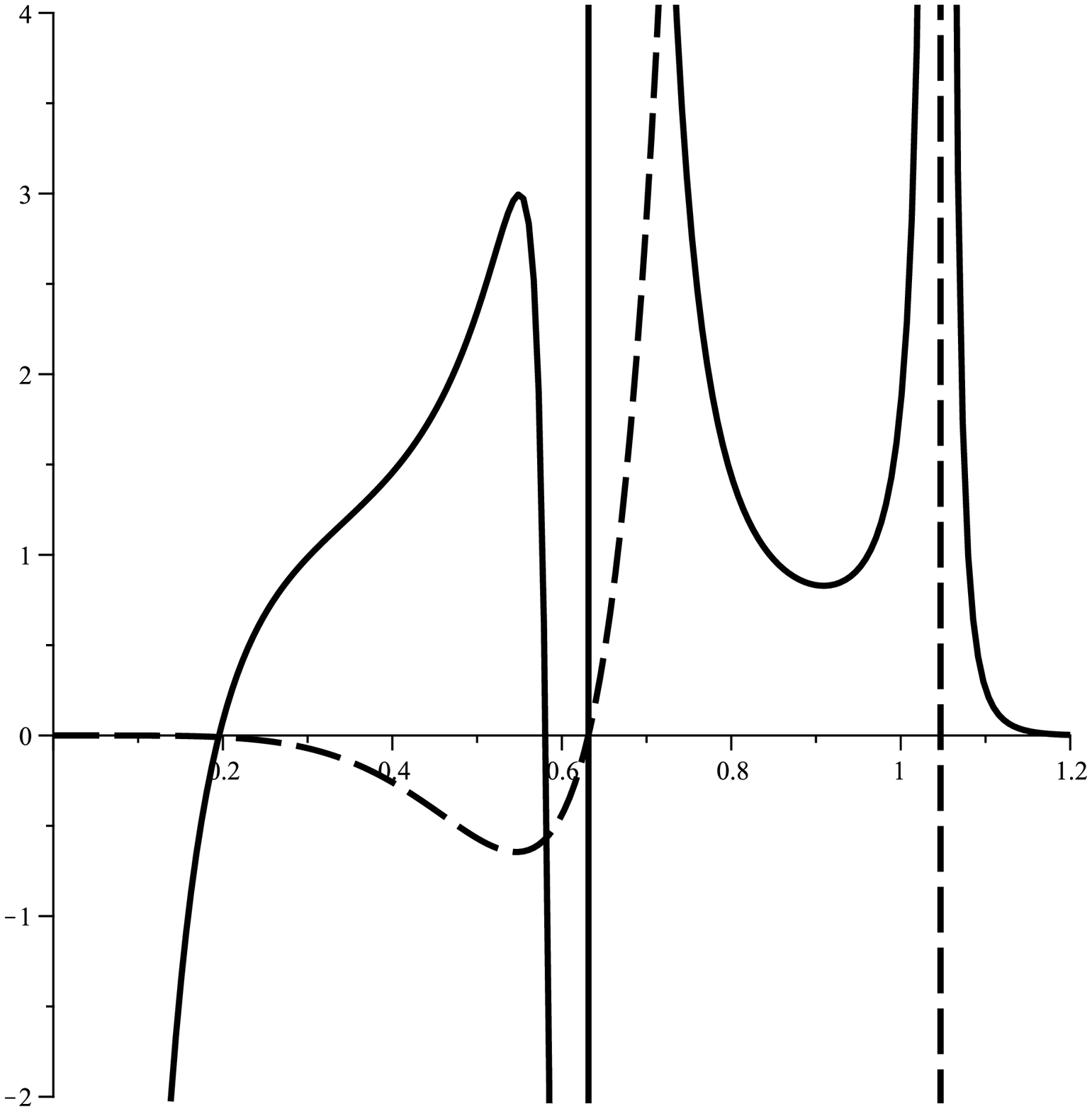} & \epsfxsize=6cm %
\epsffile{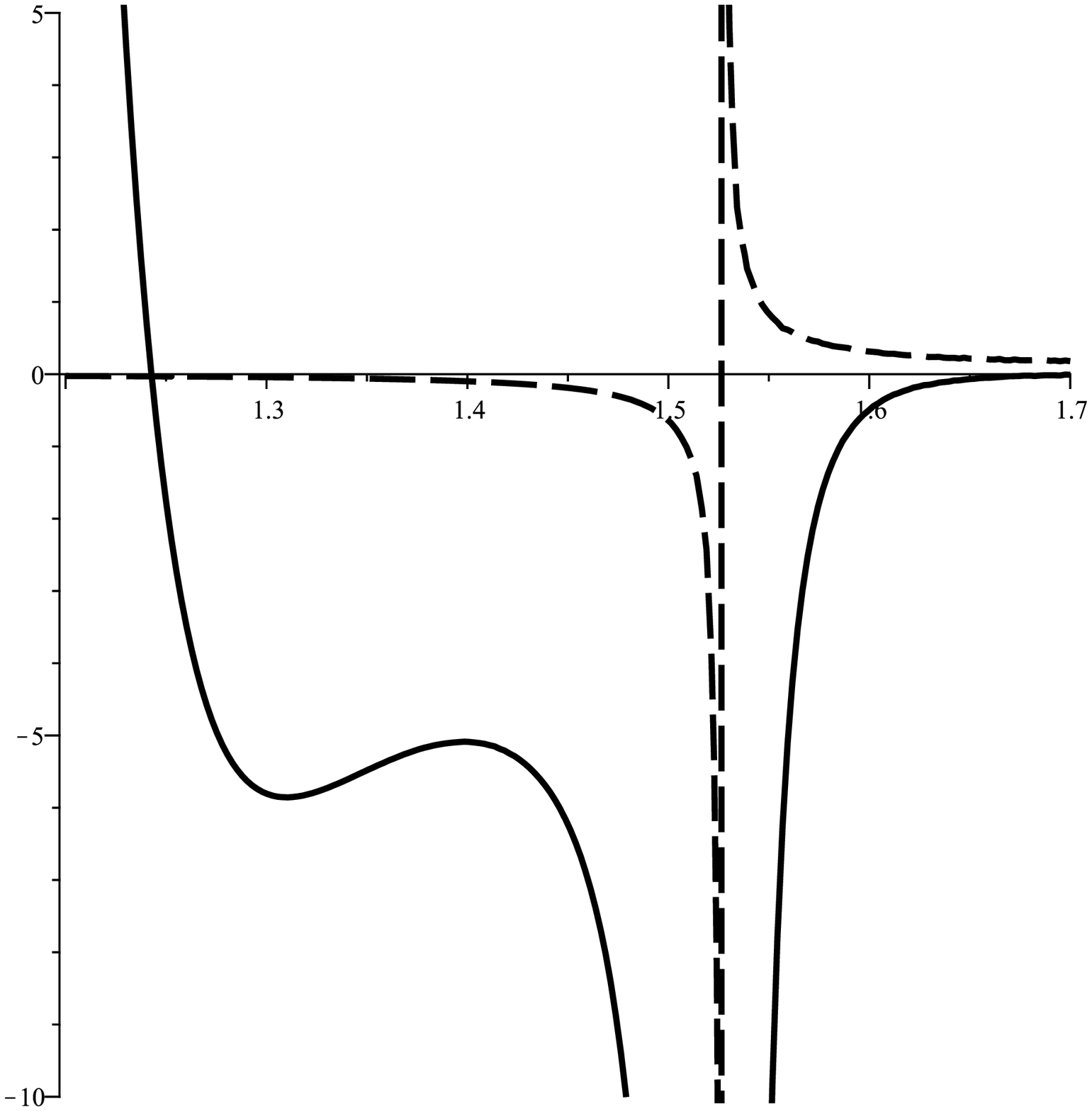} \\
\epsfxsize=6cm \epsffile{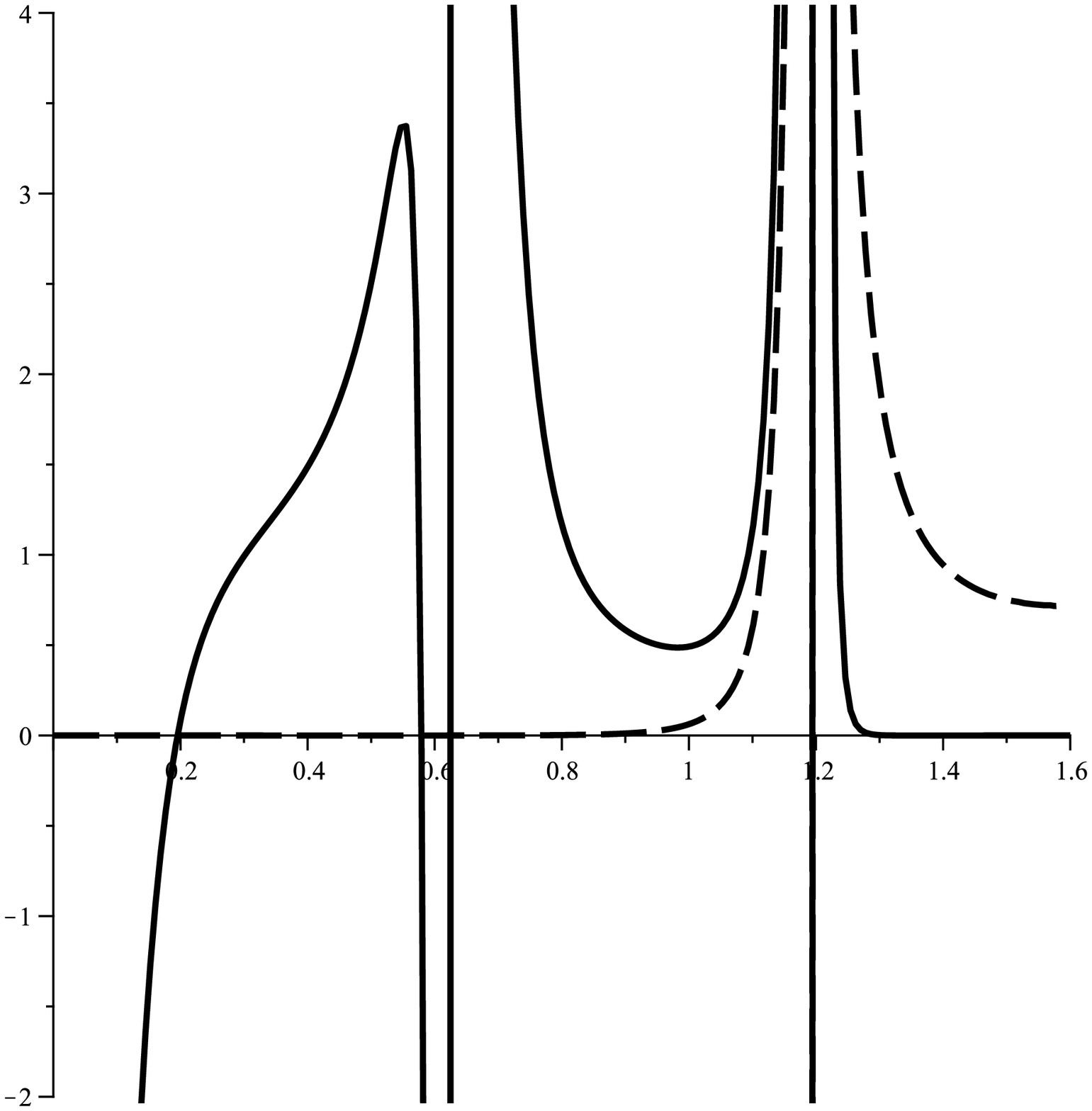} & \epsfxsize=6cm %
\epsffile{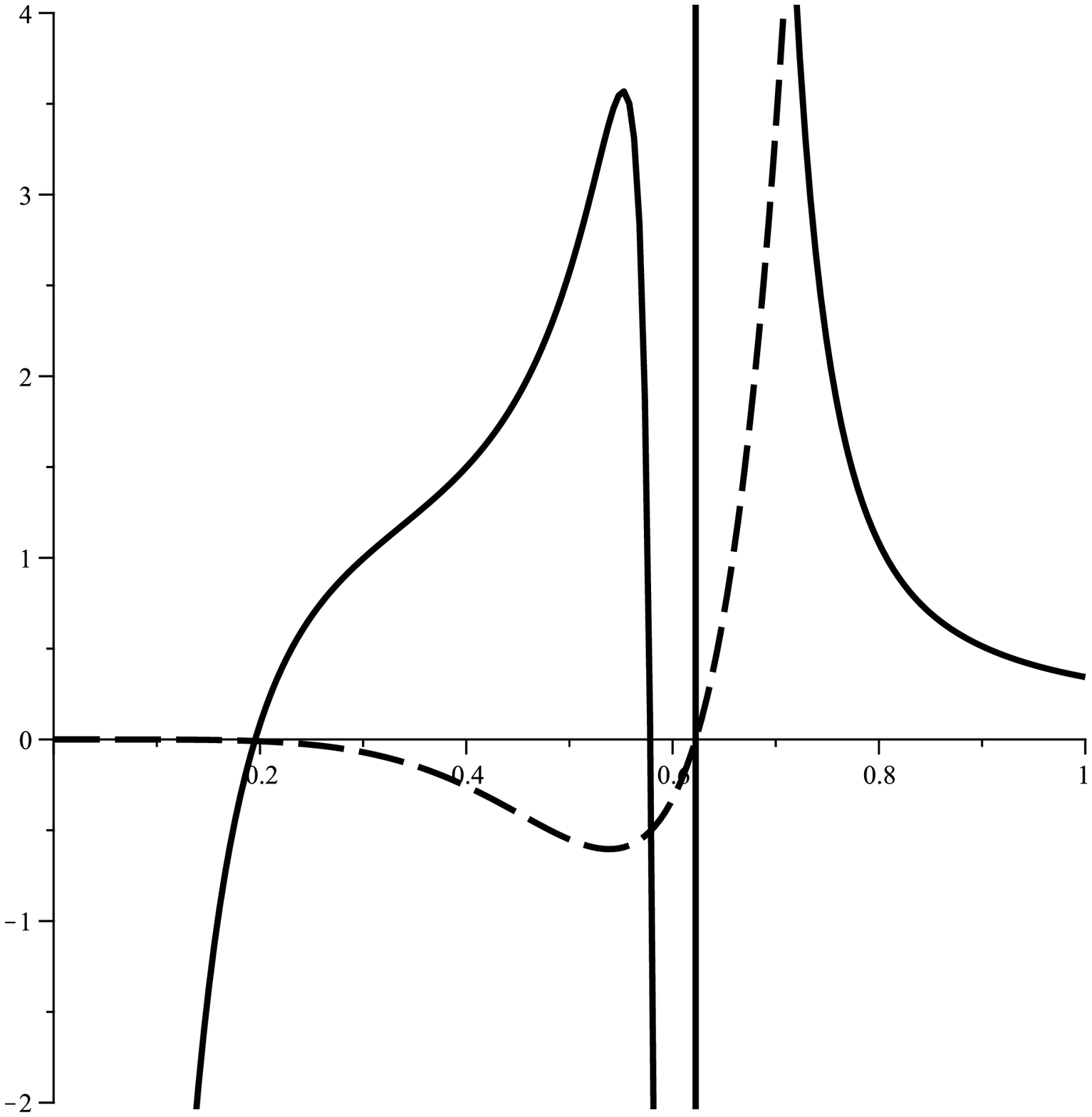}%
\end{array}
$%
\caption{\textbf{"LNEF branch:"} $\mathcal{R}$ (continuous line - Case $II$%
), $C_{Q}$ (dashed line) diagrams for $d=7$, $q=1$ and $\protect\beta =1$.
\newline
up: $P=0.75P_{c}$ for different scales \newline
down: $P=P_{c}$ left and $P=1.1P_{c}$ right \newline
.}
\label{13}
\end{figure}

As one can see, special case of existence of different orders of pressure,
the $M_{P}^{2}$ may contribute to number of the divergencies of TRS. This
contribution may lead to existence of extra divergencies for the Ricci
scalar which do not coincide with any phase transition point of the heat
capacity. In our case (obtained solutions), due to the existence of linear
function of $P$, there will be no contribution to number of divergencies of
the heat capacity. Therefore we expect that although the total behavior of
TRS may change from that of previous TRS, the places of divergencies
coincides. Plotting graphs confirm that the location of divergence points of
new TRS will be the same as those of previous TRS (see Figs. \ref{12} and %
\ref{13} and compare them with Figs. \ref{101} - \ref{112}).


\section{Critical values through heat capacity in the extended phase space}

In this section we will introduce a new method for obtaining critical
pressure and horizon radius and studying the critical behavior of the system
in the extended phase space. In last section it was pointed out that phase
transition points in extended phase space appeared as divergence points in
the heat capacity. Therefore, one can use divergencies of the heat capacity;
hence root of the denominator of the heat capacity for obtaining critical
values and studying critical behavior of the system. For more
clarifications, we explain the method step by step.

First step is devoted to finding the heat capacity for the system under
consideration. Next, we use the analogy between pressure and cosmological
constant to rewrite the obtained heat capacity in the extended phase space.
Then, we take into account denominator of the heat capacity and solve it
with respect to pressure to obtain a relation for pressure. It should be
pointed out that this relation is not same as the one that was derived for
pressure in Eq. (\ref{Pressure}). In other words, these two expressions are
independent of each other.

The newly obtained relation for pressure contains different information
regarding critical points and the critical behavior of the system. The
existence of maximum in obtained relation is representing critical pressure
and horizon in which phase transition takes place in it. The general
behavior of the pressure in case of this relation (one or several maximum
points) indicates the critical behavior of the system (one or several phase
transition, triple point, regular and non-regular phase transitions). In
other words, only by studying the diagrams of this relation for different
values of metric function parameters one can study the critical behavior of
the system. It is worthwhile to mention that absence of the maximum is
representing non-critical system.

In order to elaborate the efficiency of the mentioned method, we
conduct a study in case of black holes that we have been studied
in this paper. As it was mentioned before, one can use Eq.
(\ref{HC}) for obtaining heat capacity for these black holes.
Using denominator of the heat capacity and solving it with respect
to pressure one can find following relations for pressure for $5$
and $7$ dimensions
\begin{equation}
P=\frac{\left[ q^{2}\left( q^{2}+3\beta ^{2}r^{2d-4}\right) +2\beta
^{4}r^{4d-8}\left( 1-\Gamma ^{3}\right) \right] }{2\pi r^{2d-4}\Gamma \left(
2\beta ^{2}r^{2d-4}\left( \Gamma -1\right) +q^{2}\left( \Gamma -2\right)
\right) }\ln \left( \frac{2\beta ^{2}r^{2d-4}\left( \Gamma -1\right) }{q^{2}}%
\right) +\frac{\Theta}{\mathcal{W}} \label{P}
\end{equation}
where
\begin{equation}
\Theta=\left\{
\begin{array}{cc}
\mathcal{A}_{1}+\mathcal{A}_{2}+\mathcal{A}_{3} & d=5 \\
\mathcal{B}_{1}+\mathcal{B}_{2}+\mathcal{B}_{3} & d=7%
\end{array}%
\right. ,
\end{equation}
in which $\mathcal{A}_{i}$'s and $\mathcal{B}_{i}$'s are
\begin{eqnarray}
&&\mathcal{A}_{1}=\frac{7}{32}\Gamma ^{2}\beta ^{3}r^{8}\left(
q^{2}+2\beta
^{2}r^{6}\left( 1-\Gamma \right) \right) \left[ q^{4}-2\beta ^{4}r^{12}-%
\frac{11}{2}q^{2}\beta ^{2}r^{6}\right] , \nonumber\\
&&\mathcal{A}_{2}=\frac{9\Gamma ^{3}\beta ^{3}r^{8}}{32}\left[
\frac{92\beta ^{6}r^{18}\left( 1-\Gamma \right) }{9}+\frac{4\beta
^{4}r^{16}\left(
1-\Gamma \right) }{3}-21q^{2}\beta ^{4}r^{12}\left( \Gamma -\frac{235}{189}%
\right) \right] , \nonumber\\
&&\mathcal{A}_{3}=\frac{9\Gamma ^{3}\beta ^{3}r^{8}}{32}\left[
q^{6}-2q^{2}\beta ^{2}r^{10}\left( \Gamma -\frac{4}{3}\right) -\frac{%
34q^{4}\beta ^{2}r^{6}\left( \Gamma -\frac{241}{68}\right) }{9}-\frac{%
2q^{4}r^{4}\left( \Gamma -2\right) }{3}\right] , \nonumber
\\ \nonumber\\ \nonumber\\&&\mathcal{B}_{1}=\frac{11}{36}\Gamma ^{2}\beta
^{3}r^{12}\left( q^{2}+2\beta ^{2}r^{10}\left( 1-\Gamma \right)
\right) \left[ q^{4}-\frac{3}{2}\beta
^{4}r^{20}-\frac{27}{4}q^{2}\beta ^{2}r^{10}\right] ,
\nonumber\\
&&\mathcal{B}_{2}=\frac{25\Gamma ^{3}\beta ^{3}r^{12}}{36}\left[ \frac{%
177\beta ^{6}r^{30}\left( 1-\Gamma \right) }{25}+\frac{9\beta
^{4}r^{28}\left( 1-\Gamma \right) }{5}-\frac{801}{50}q^{2}\beta
^{4}r^{20}\left( \Gamma -\frac{326}{267}\right) \right] ,
\nonumber\\
&&\mathcal{B}_{3}=\frac{25\Gamma ^{3}\beta ^{3}r^{12}}{36}\left[ q^{6}-\frac{27%
}{10}q^{2}\beta ^{2}r^{18}\left( \Gamma -\frac{4}{3}\right) -\frac{%
86q^{4}\beta ^{2}r^{10}\left( \Gamma -\frac{1073}{344}\right) }{25}-\frac{%
9q^{4}r^{8}\left( \Gamma -2\right) }{10}\right] ,\nonumber
\end{eqnarray}%
and $\mathcal{W}=-\frac{1}{2}\pi \Gamma ^{4}\beta ^{5}r^{20}\left[
2\beta ^{2}r^{6}\left( \Gamma -1\right) +q^{2}\left( \Gamma
-2\right) \right] $.

\begin{figure}[tbp]
$%
\begin{array}{cc}
\epsfxsize=6cm \epsffile{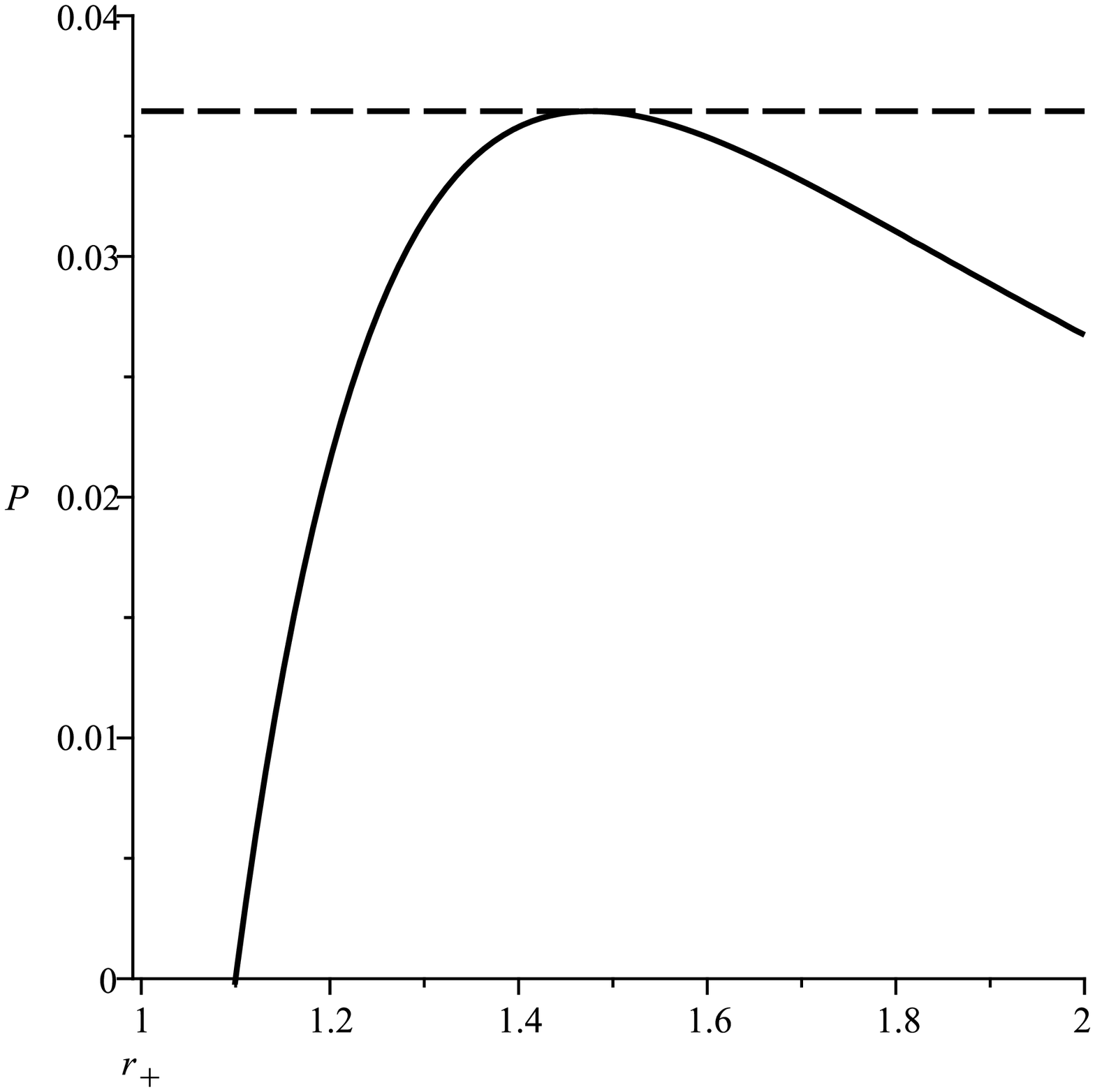} & \epsfxsize=6cm \epsffile{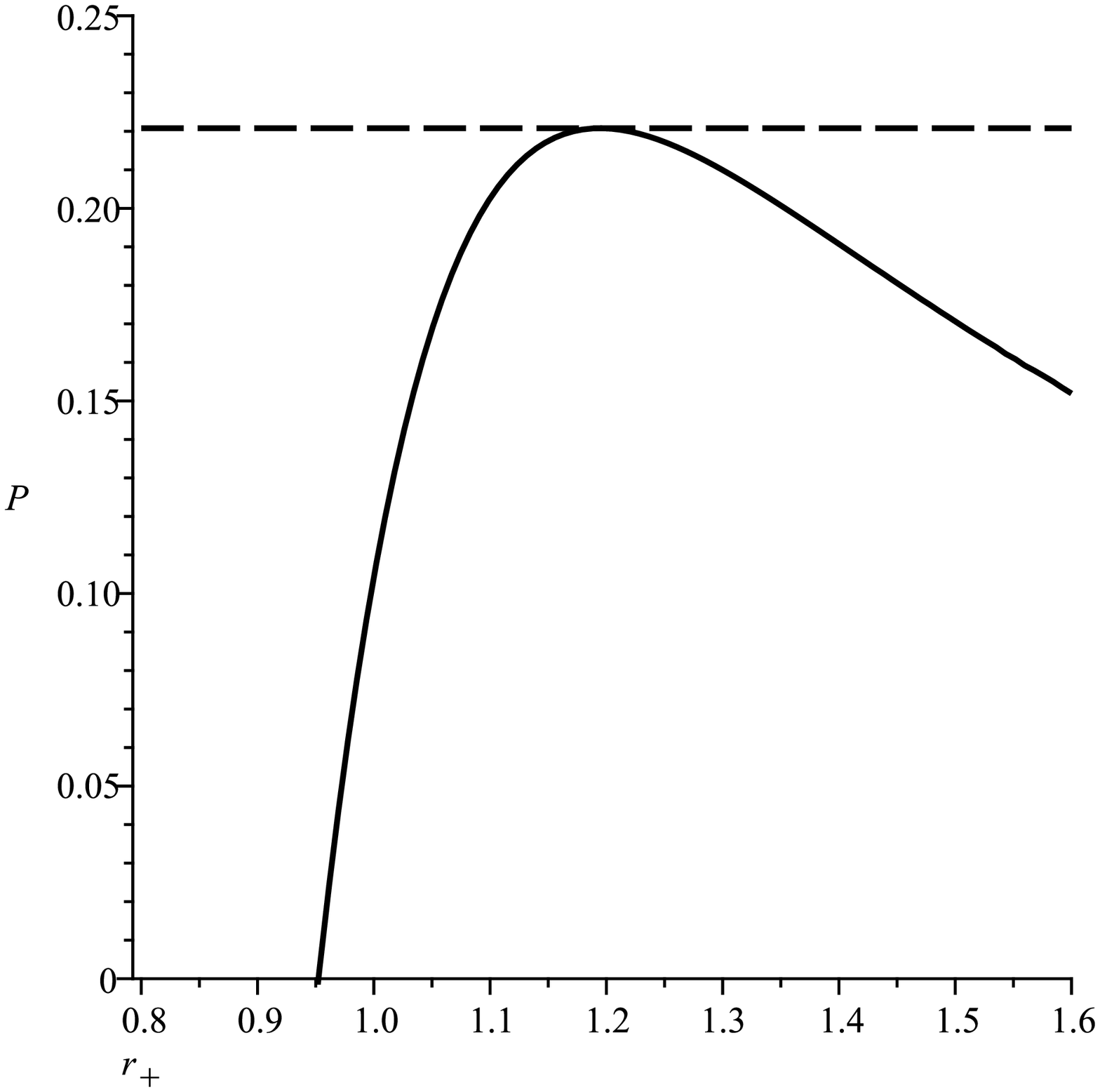}%
\end{array}
$%
\caption{\textbf{"LNEF branch:"} $P$ versus $r_{Q+}$ diagrams for $q=1$ and $%
\protect\beta =1$. \newline
left panel: $d=5$(continues line) and $P=0.0360$ (dashed line); right panel:
$d=7$, (continues line) and $P=0.2207$ (dashed line). }
\label{14}
\end{figure}

Next, by considering the values that are considered for different
parameters in previous sections (the ones in tables $1$ and $2$),
we plot following diagrams for $5$ and $7$ dimensional cases (Fig.
\ref{14}).

It is evident that for considering the mentioned values one finds that there
is a maximum for plotted graphs of pressure versus horizon radius. As one
can see, the pressure and horizon radius of this maximum are exactly what we
have obtained in tables 1 and 2 for $5$ and $7$ dimensional cases. On the
other hand, we see that the behavior of the pressure is exactly what we are
expecting in thermodynamical aspect. There is a critical value for pressure
(the maximum) in which for smaller values of the that critical pressure,
there exists two critical pressure and for pressures larger than the
critical pressure (maximum) there is no value for pressure, therefore, no
phase transition will take place. This behavior is consistent with what we
have observed in studying $T-v$ diagrams.

There are several results regarding this study that we highlight them:

1) The thermodynamical behavior of the system, which in our case is black
hole, is consistent in context of all three approaches that we have used to
study the critical behavior of the system. Therefore, all of these
approaches yield consisting results.

2) In order to study the thermodynamical behavior of the system, and
existence of regular and non-regular critical behavior, studying the
denominator of the heat capacity is sufficient. In other words, solving the
denominator with respect to pressure, leads to a relation in which maximums
are presenting critical values and phase transition points. Therefore,
instead of using Eqs. \ref{cr1} and \ref{cr2}, one can study the denominator
of the heat capacity and its behavior.

3) Equations. \ref{cr1} and \ref{cr2} are usually presenting complicated set
of equations which may be difficult to solve for obtaining critical values
whereas studying the maximums of the obtained relation from denominator of
the heat capacity for pressure is much more easier to deal with.

\section{Conclusions}

In this paper, we have considered two classes of nonlinear electrodynamics
in presence of Einstein gravity and studied their phase transitions. At
first, by considering cosmological constant as thermodynamical pressure and
its conjugating variable as volume we have extended the phase space and
regarded the interpretation of total mass of black hole as the Enthalpy.
Studying calculated critical values through three different types of phase
diagrams resulted into phase transition taking place in the critical values.
$P-v$, $T-v$ and $G-T$ diagrams representing similar behavior near critical
points similar to their corresponding diagrams in Van der Waals liquid-gas
system.

Studying the effects of nonlinearity parameter on the phase diagrams
revealed the fact that as nonlinearity parameter increases the critical
temperature and pressure decrease which indicates that for large values of
nonlinearity parameter the system needs less energy (mass) absorption to
have phase transition. Due to fact that for large $\beta$ these nonlinear
models reduce to Maxwell theory, one might conclude that the lowest
temperature and pressure in which phase transition takes place, belongs to
Maxwell theory. Besides, we found that the universal ratio of $\frac{%
P_{c}v_{c}}{T_{c}}$ increases as nonlinearity parameter increases. It is
worthwhile to mention that subcritical isobars in $T-v$ are decreasing
functions of nonlinearity parameter. These subcritical isobars are
representing the region in which phase transition takes place which its
length is a decreasing function of nonlinearity parameter.

Next, studying the effects of dimensionality showed that, for higher
dimensional black holes, phase transitions take place in higher temperature
and the difference between Gibbs free energy of different phases grows
larger. We found that as dimensionality of black holes increases, phase
transition for obtaining stable state becomes more difficult and black holes
need to absorb more mass in order to have phase transition. It is worthwhile
to mention the fact that dimensionality also increases the critical
temperature and pressure drastically whereas it decreases the critical
horizon. In other words, in comparing critical pressure and temperature of
several dimensions, the highest values of these critical quantities belong
to the highest dimensions.

One should take this fact into account that for small values of
nonlinearity, changes in critical values are greater comparing to large
values of nonlinearity parameter. This fact is evident from studying phase
diagrams and tables. These differences in critical values are due to fact
that as nonlinearity parameter decreases to values near zero the power of
nonlinear electromagnetic fields grows stronger. This behavior indicates
that as nonlinear electromagnetic field grows stronger, for having phase
transition, the black hole needs to absorb more mass; hence the total mass
of black hole must increase in order to have small/large (unstable/stable)
black hole phase transition. In other words, in generalizing from linear
theory of electrodynamics to nonlinear ones, we have fundamentally modifying
the interior thermodynamical structure of the black holes. It is very
important to understand the effects of the nonlinearity power. As we have
seen the power of the nonlinearity decreases as the nonlinearity parameter
increases which caused the critical behavior of the system be Maxwell like.
Considering the mentioned nonlinear electrodynamics, we have shown not only
the metric function and the electromagnetic field will be generalized, but
also the critical behavior of the system will be modified in such a way that
in case of large $\beta$, the system recover the behaviors of
Reissner--Nordstr\"{o}m black holes. On the other sides, for small values of
nonlinearity parameter, the power of nonlinearity grows drastically and the
critical values were highly modified. It is worthwhile to mention that for $%
\beta \longrightarrow 0$, the electromagnetic field vanishes and the system
switch to the Schwarzschild--like behavior which we are expecting to see no
phase transition. Taking this behavior into account, one might argue that
before system attaining Schwarzschild--like behavior, it will have the
highest critical temperature and pressure and the lowest critical horizon.
In other words, near the Schwarzschild--like limit, critical horizon will be
so small and in Schwarzschild case, the critical horizon will vanish and
therefore there will be no phase transition.

Next, we employed heat capacity and GTD method in order to study the
obtained critical points in extended phase space. It was pointed out that
phase transition points of the extended phase space only appeared as
divergencies of the heat capacity which were denoted as type two phase
transitions. The characteristic behavior of heat capacity (number of
divergencies of heat capacity) was similar to the one in study conducted in
context of extended phase space specially $T-v$ diagrams. On the other hand,
two new geometrothermodynamical metrics were introduced for studying
critical behavior of the system. In these cases, the divergencies of TRS of
thermodynamical metrics coincided with phase transitions of the heat
capacity. Also, the characteristic behavior of TRS for different cases of
the critical values was similar to the one in extended phase space.
Therefore, we established two new methods of GTD for studying critical
behavior of the system in context of extended phase space.

Finally, we showed that one can use the denominator of the heat capacity for
obtaining the critical pressure, horizon radius and the characteristic
behavior of the system near critical point (regular and non-regular ones).
In other words, solving denominator of the heat capacity with respect to
pressure lead to a relation in which maximums were representing phase
transition points of the system. Using the denominator of the heat capacity
has certain advantages that make it more favorable in comparison with other
approaches. Also, we showed that results that were found using GTD, heat
capacity and extended phase space lead to consisting results. In other
words, these three pictures have consisting machineries for describing
critical behavior of the system.

For the future works, one may generalize obtained static solutions to a case
of dynamical ones. It means that one may consider nonlinearly dynamical
black hole solutions to investigate Hawking radiation. We know that the
Hawking radiation decreases the total mass of black hole and according to
our obtained results, one may say that this radiation shifts black hole from
stable to unstable phase. In other words, Hawking radiation mechanism (black
hole evaporation) affects the stability and phase transition of black holes
fundamentally and makes black holes unstable in our models. It will be
constructive to study a time dependent (dynamical) spacetime through Hawking
radiations near the critical point.

\begin{acknowledgements}
We thank the Shiraz University Research Council. This work has
been supported financially by the Research Institute for Astronomy
and Astrophysics of Maragha, Iran.
\end{acknowledgements}


\begin{thebibliography}{99}
\bibitem{AdS/CFT1} J. Maldacena, Adv. Theor. Math. Phys. \textbf{2}, 231
(1998).

\bibitem{AdS/CFT2} E. Witten, Adv. Theor. Math. Phys. \textbf{2}, 253 (1998).

\bibitem{AdS/CFT3} S. S. Gubser, I. R. Klebanov, and A. M. Polyakov, Phys.
Lett. B \textbf{428}, 105 (1998).

\bibitem{AdS/CFT4} O. Aharony, S. S. Gubser, J. Maldacena, H. Ooguri, and Y.
Oz, Phys. Rept. \textbf{323}, 183 (2000).

\bibitem{VariableLambda1} J. Creighton and R. B. Mann, Phys. Rev. D \textbf{%
52}, 4569 (1995).

\bibitem{VariableLambda2} G. W. Gibbons, R. Kallosh, and B. Kol, Moduli,
Phys. Rev. Lett. \textbf{77} 4992 (1996).

\bibitem{PressureLambda1} D. Kastor, S. Ray and J. Traschen, Class. Quantum
Gravit. \textbf{26}, 195011 (2009).

\bibitem{PressureLambda2} M. Cvetic, G. W. Gibbons, D. Kubiznak and C. N.
Pope, Phys. Rev. D \textbf{84}, 024037 (2011).

\bibitem{PressureLambda3} B. P. Dolan, Class. Quantum Gravit. \textbf{28},
125020 (2011).

\bibitem{PressureLambda4} B. P. Dolan, Class. Quantum Gravit. \textbf{28},
235017 (2011).

\bibitem{PressureLambda5} D. Kubiznak and R. B. Mann, JHEP \textbf{07}, 033
(2012).

\bibitem{PressureLambda6} S. Gunasekaran, R. B. Mann and D. Kubiznak, JHEP
\textbf{11}, 110 (2012).

\bibitem{PressureLambda7} R. Banerjee and D. Roychowdhury, Phys. Rev. D
\textbf{85}, 104043 (2012).

\bibitem{PressureLambda8} A. Lala and D. Roychowdhury, Phys. Rev. D \textbf{%
86}, 084027 (2012).

\bibitem{PressureLambda9} R. Banerjee and D. Roychowdhury, Phys. Rev. D
\textbf{85}, 044040 (2012).

\bibitem{PressureLambda10} S. H. Hendi and M. H. Vahidinia, Phys. Rev. D
\textbf{88}, 084045 (2013).

\bibitem{PressureLambda11} D. C. Zou, S. J. Zhang and B. Wang, Phys. Rev. D
\textbf{89}, 044002 (2014).

\bibitem{HawkingPage} S. W. Hawking and D. N. Page, Commun. Math. Phys.
\textbf{87}, 577 (1983).

\bibitem{Witten} E. Witten, Adv. Theor. Math. Phys. \textbf{2}, 505 (1998).

\bibitem{delph1} D. H. Delphenich, [arXiv:0309108].

\bibitem{delph2} D. H. Delphenich, [arXiv:0610088].

\bibitem{Hassaine1} M. Hassaine and C. Martinez, Phys. Rev. D \textbf{75},
027502 (2007).

\bibitem{Hassaine2} S. H. Hendi and H. R. Rastegar-Sedehi, Gen. Relativ.
Gravit. \textbf{41}, 1355 (2009).

\bibitem{Hassaine3} S. H. Hendi, Phys. Lett. B \textbf{677}, 123 (2009).

\bibitem{Hassaine4} H. Maeda, M. Hassaine and C. Martinez, Phys. Rev. D
\textbf{79}, 044012 (2009).

\bibitem{Hassaine5} S. H. Hendi and B. Eslam Panah, Phys. Lett. B \textbf{684%
}, 77 (2010).

\bibitem{Hassaine6} S. H. Hendi, Eur. Phys. J. C \textbf{69}, 281 (2010).

\bibitem{Oliveira1} H. P. de Oliveira, Class. Quantum Gravit. \textbf{11},
1469 (1994).

\bibitem{Oliveira2} B. L. Altshuler, Class. Quantum Gravit. \textbf{7}, 189
(1990).

\bibitem{Soleng} H. H. Soleng, Phys. Rev. D \textbf{52}, 6178 (1995).

\bibitem{HendiJHEP} S. H. Hendi, JHEP \textbf{03}, 065 (2012).

\bibitem{Seiberg} N. Seiberg and E. Witten, JHEP \textbf{09}, 032 (1999).

\bibitem{Born-Lnfeld} M. Born and L. Infeld, Proc. R. Soc. A \textbf{144},
425 (1934).

\bibitem{Fradkin1} E. Fradkin and A. Tseytlin, Phys. Lett. B \textbf{163},
123 (1985).

\bibitem{Fradkin2} R. Matsaev, M. Rahmanov and A. Tseytlin, Phys. Lett. B
\textbf{193}, 207 (1987).

\bibitem{Fradkin3} E. Bergshoeff, E. Sezgin, C. Pope and P. Townsend, Phys.
Lett. B \textbf{188}, 70 (1987).

\bibitem{Fradkin4} C. Callan, C. Lovelace, C. Nappi and S. Yost, Nucl. Phys.
B \textbf{308}, 221 (1988).

\bibitem{Fradkin5} O. Andreev and A. Tseytlin, Nucl. Phys. B \textbf{311},
221 (1988)

\bibitem{Fradkin6} R. Leigh, Mod. Phys. Lett. A \textbf{04}, 2767 (1989).

\bibitem{Weinhold} F. Weinhold, J. Chem. Phys. \textbf{63}, 2479 (1975).

\bibitem{Ruppeiner} G. Ruppeiner, Phys. Rev. A \textbf{20}, 1608 (1979).

\bibitem{Janyszek1986} H. Janyszek, Rep. Math. Phys. \textbf{24}, 1 (1986).

\bibitem{Brody} E. J. Brody, Phys. Rev. Lett. \textbf{58}, 179 (1987).

\bibitem{Dolan2002} B. P. Dolan, D. A. Johnston and R. Kenna, J. Phys. A
\textbf{35}, 9025 (2002).

\bibitem{Janke2004} W. Janke, D. A. Johnston and R. Kenna, Physica A \textbf{%
336}, 181 (2004).

\bibitem{Ferrara} S. Ferrara, G. W. Gibbons and R. Kallosh, Nucl. Phys. B
\textbf{500}, 75 (1997).

\bibitem{Cai1999} R. G. Cai and J. H. Cho, Phys. Rev. D \textbf{60}, 067502
(1999).

\bibitem{Aman2003} J. E. Aman, I. Bengtsson and N. Pidokrajt, Gen. Relativ.
Gravit. \textbf{35}, 1733 (2003) .

\bibitem{Carlip} S. Carlip and S. Vaidya, Class. Quantum Gravit. \textbf{20}%
, 3827 (2003).

\bibitem{Mirza} B. Mirza and M. Zamani-Nasab, JHEP \textbf{06}, 059 (2007) .

\bibitem{HPEM} S. H. Hendi, S. Panahiyan, B. Eslam Panah and M. Momennia,
\textit{A new approach toward geometrical concept of black hole
thermodynamics}, submitted for publication.

\bibitem{Quevedo2007} H. Quevedo, J. Math. Phys. \textbf{48}, 013506 (2007).

\bibitem{Quevedo2008} H. Quevedo and A. Sanchez, JHEP \textbf{09}, 034
(2008).

\bibitem{Mo} J. X. Mo, X. X. Zeng, G. Q. Li, X. Jiang and W. B. Liu, JHEP
\textbf{10}, 056 (2013).

\bibitem{Zhang} J. L. Zhang, R. G. Cai and H. Yu, JHEP \textbf{02}, 143
(2015).

\bibitem{QuevedoP2011} H. Quevedo, A. Sanchez, S. Taj and A. Vazquez, Gen.
Relativ. Gravit. \textbf{43}, 1153 (2011).

\bibitem{Han} Y. Han, G. Chen, Phys. Lett. B \textbf{714}, 127 (2012).

\bibitem{Bravetti} A. Bravetti, D. Momeni, R. Myrzakulov and A. Altaibayeva,
Adv. High Energy Phys. \textbf{2013}, 549808 (2013).

\bibitem{HendiAnn2} S. H. Hendi, Ann. Phys. \textbf{346}, 42 (2014).

\bibitem{HendiAnn1} S. H. Hendi, Ann. Phys. \textbf{333}, 282 (2013).

\bibitem{Hendiothers1} S. H. Hendi and M. Allahverdizadeh, Adv. High Energy
Phys. 390101 (2014).

\bibitem{Hendiothers2} S. H. Hendi and A. Sheykhi, Phys. Rev. D \textbf{88},
044044 (2014).

\bibitem{Brewin} L. Brewin, Gen. Relativ. Gravit. \textbf{39}, 521 (2007).

\bibitem{SmarrNew1} D. Kastor, S. Ray and J. Traschen, Class. Quant. Grav.
\textbf{27}, 235014 (2010).

\bibitem{SmarrNew2} R. G. Cai, L. M. Cao, L. Li and R.Q. Yang, JHEP \textbf{%
09}, 005 (2013).

\bibitem{SmarrNew3} D. C. Zou, S. J. Zhang and B. Wang, Phys. Rev. D \textbf{%
89}, 044002 (2014).

\bibitem{SmarrNew4} Z. Sherkatghanad, B. Mirza, Z. Mirzaeyan and S. A. H.
Mansoori, [arXiv:1412.5028].
\end{thebibliography}
\end{document}